\documentclass[10pt,journal,compsoc]{IEEEtran}

\ifCLASSOPTIONcompsoc
  \usepackage[nocompress]{cite}
\else
  \usepackage{cite}
\fi

\usepackage{tikz}
\usepackage{amsmath}
\usepackage{url}
\usepackage{hyperref}

\usepackage{filecontents}
\usepackage{graphicx}
\usepackage{hhline}
\usepackage{float}
\usepackage{xcolor}
\usepackage{caption}
\usepackage{subfig}
\usepackage{listings}
\usepackage{multirow}
\usepackage{enumitem}
\usepackage{tcolorbox}
\usepackage{booktabs}
\usepackage{inconsolata}
\usepackage{algorithm}
\usepackage{algorithmic}
\usepackage{mathtools}
\usepackage{amsthm}
\usepackage{amssymb}
\usepackage{mdframed}
\usepackage{tablefootnote}
\usepackage{refcount}
\usepackage{courier}
\usepackage[normalem]{ulem}

\usepackage{mathtools}

\usepackage{tikz}
\newcommand*\circled[1]{\tikz[baseline=(char.base)]{
            \node[shape=circle,draw,inner sep=0.5pt] (char) {#1};}}

\definecolor{light-gray}{gray}{0.95}
\DeclareMathAlphabet{\mathcal}{OMS}{cmsy}{m}{n}

\newcommand{\code}[1]{\colorbox{light-gray}{\texttt{#1}}}
\renewcommand{\P}{\mathcal{P}}

\newtheorem*{assumption*}{\assumptionnumber}
\providecommand{\assumptionnumber}{}


\newtheorem{definition}{Definition}

\begin{document}

\title{Tokoin: A Coin-Based Accountable Access Control Scheme for Internet of Things}
\author{Chunchi~Liu,
        Minghui~Xu,
        Hechuan~Guo,
        Xiuzhen~Cheng,~\IEEEmembership{Fellow,~IEEE,}
        Yinhao~Xiao,
        Dongxiao~Yu,
        Bei~Gong,
        Arkady~Yerukhimovich,
        Shengling~Wang,
        and~Weifeng~Lv
        
   \IEEEcompsocitemizethanks{
	
	\IEEEcompsocthanksitem C. Liu, M. Xu (corresponding author), X. Cheng, and A. Yerukhimovich are with Computer Science Department, The George Washington University. E-mail: \{liuchunchi, mhxu, cheng, arkady\}@gwu.edu \hfil\break 
	\IEEEcompsocthanksitem  H. Guo and D. Yu are with School of Computer Science and Technology, Shandong University. E-mail: \{ghc, dxyu\}@sdu.edu.cn \hfil\break 
	\IEEEcompsocthanksitem Y. Xiao is with Guangdong University of Finance and Economics. E-mail: xyh3984@gmail.com \hfil\break 
	\IEEEcompsocthanksitem B. Gong is with Beijing University of Technology. E-mail: tekkman\_blade@126.com \hfil\break 
	\IEEEcompsocthanksitem S. Wang is with Beijing Normal University. E-mail: wangshengling@bnu.edu.cn \hfil\break 
	\IEEEcompsocthanksitem W. Lv is with Beihang University. E-mail: lwf@nlsde.buaa.edu.cn \hfil\break 
%
	}
}

%
%
%
%
%
%
%

\markboth{Journal of \LaTeX\ Class Files,~Vol.~14, No.~8, August~2015}%
{Shell \MakeLowercase{\textit{et al.}}: Bare Demo of IEEEtran.cls for Computer Society Journals}

\IEEEtitleabstractindextext{%

\begin{abstract}
With the prevalence of Internet of Things (IoT) applications, IoT devices interact closely with our surrounding environments, bringing us unparalleled smartness and convenience. However, the development of secure IoT solutions is getting a long way lagged behind, making us exposed to common unauthorized accesses that may bring malicious attacks and unprecedented danger to our daily life.  \textbf{Overprivilege attack}, a widely reported phenomenon in IoT that accesses unauthorized or excessive resources, is notoriously hard to prevent, trace and mitigate.
To tackle this challenge, we propose Tokoin-Based Access Control (TBAC), an accountable access control model enabled by blockchain and Trusted Execution Environment (TEE) technologies, to offer fine-graininess, strong auditability, and access procedure control for IoT. TBAC materializes the virtual access power into a definite-amount and secure  cryptographic coin termed ``tokoin'' (token+coin), and manages it using atomic and accountable state-transition functions in a blockchain. 
A tokoin can be created only by the resource owner, and is programmed with a fine-grained access policy defining ``{\bf who} is allowed to do {\bf what} at {\bf when} in {\bf where} by {\bf how}''. Such a policy specifies the access requirements  (the \emph{4W}) to be satisfied for granting the access, and access behavioral constraints (the \emph{1H}) that must be strictly followed during the resource access. To the best of our knowledge, we are the first to realize such an \textit{access procedure control} that governs actual access operations throughout the whole access process. The strong-auditability is achieved with blockchain and a TEE-enabled trusted access control object such that trust can be extended from on-chain to off-chain making all access activities securely monitored and auditable. The tokoin is peer-to-peer transferable, and can be modified only by the resource owner when necessary, realizing a dynamic and flexible access control that can accommodate the ever-changing physical world, without relaxing any security requirement. 
We fully implement TBAC with well-studied cryptographic primitives and blockchain platforms and present a readily available APP for regular users. We also present a case study to demonstrate how TBAC is employed to enable autonomous in-home cargo delivery while guaranteeing the access policy compliance and home owner's physical security by regulating the physical behaviors of the deliveryman. 


\end{abstract}

}

\maketitle

\IEEEdisplaynontitleabstractindextext

\IEEEpeerreviewmaketitle

\IEEEraisesectionheading{\section{Introduction}\label{sec:introduction}}

\IEEEPARstart{W}{ith} the rapid development of the Internet of Things (IoT), IoT devices have become much smaller, smarter, and powerful than ever before.  Iconic products such as Google Home, Amazon Alexa, and Samsung SmartThings have brought great convenience to human life. According to \cite{StatisticaSmartHome}, household penetration of smart home will reach 36.6\% by the end of 2020 and is expected to hit 57.2\% by 2025. 
Unfortunately, it has been widely reported by many researchers \cite{kumar2018skill}\cite{fernandes2016security} and major public media \cite{washingtonpostalexa}\cite{usatodayalexa}\cite{cnngoogle} that a variety of mainstream IoT devices, including the ones mentioned above, have been secretly accessed without authorization, 
which greatly undermines the security and privacy of an ordinary user. 

Unauthorized access attacks can be ascribed as \textbf{overprivilege attacks} that aim to access unauthorized or excessive resources in stealth, making it hard to trace what have happened and who are responsible for the bad. Such attacks are very common in practice and effectively defending them is a grand challenge in IoT security. 
They are mainly caused by three significant design deficiencies of current IoT access control schemes:  \textbf{coarse granularity} of the access policy, \textbf{weak auditability} towards all access activities, and \textbf{lack of access procedure control}. Coarse granularity refers to a low expressiveness of access policy -- an inability to precisely define the required conditions for granting the access and the procedure the access must follow. Weak auditability, as the name implies, is an inability to securely log all access activities in detail. Lack of access procedure control refers to the missing of continuously examining the behavioral obedience to the pre-defined access policy during resource access. This is a systematic neglect in current access control schemes, where the action of resource access is narrowed down to a decision-making process that only relies on the initial, static access conditions at the moment when the access is requested, ignoring the access procedure that actually manipulates different resources with different access rights at different instant of time. 
Coarse granularity is the direct cause of overprivilege attacks, weak auditability implies the impotence of holding the attackers accountable, while the lack of access procedure control results in oversights to such attacks in action.  Many IoT access control mechanisms have been proposed in recent years, but unfortunately almost all of them suffer from one or more of the deficiencies mentioned above \cite{xiao2018EdgeSecurityPIEEE}.

The design objective of our Tokoin-Based Access Control mechanism (TBAC) is to provide fine-grained access control for IoT applications that can not only verify the conditions for granting the access rights but also regulate the access procedure ensuring that the access policy is strictly followed in whole, with all activities logged for auditing purpose. In TBAC, an access power is created, stored, transferred, revoked, and redeemed in the form of an accountable digital asset, namely a tokoin (token+coin), which is managed securely and auditably as a cryptographic coin with the assistance of blockchain and Trusted Execution Environment (TEE) technologies. Any user can issue tokoins representing access powers to its own resources, and any access activity must be granted by redeeming a tokoin, which relies on a TEE-enabled trusted access control object to objectively verify the access conditions and monitor the whole access procedure. A tokoin carries a fine-grained access policy specified by the resource owner to define the access conditions to be met and the access procedure to be followed. We detail TBAC in this paper by making the following contributions.


First, by combining token and coins together, we materialize the ``virtual'' access right into a definite-amount, cryptographically-secure and accountable digital asset that is a tokoin. A tokoin can only be created and revoked by the resource owner, who also has the exclusive right to define and dynamically modify the access policy of its resource. By this way the resource owner can take full control of the access to its resource down to each access activity and behavior, without the need of delegating or relying on any untrustworthy third party such as a server. Note that a tokoin itself is capable of peer-to-peer delegation and re-distribution, permitting increased flexibility. 

Second, a tokoin carries the so-called \emph{4W1H} access policy specified by the resource owner, which defines the access conditions to be met and the access procedure to be followed. The \emph{4W1H} stands for \emph{\textbf{who} is allowed to do \textbf{what} at \textbf{when} in \textbf{where} by \textbf{how}}, where the \emph{4W} present the access conditions that must be satisfied in order for an access request to be granted, and the \emph{1H} describes the access procedure that must be strictly followed. This \emph{4W1H} policy model provides fine-grained access control as it can not only precisely describe the access conditions but also capture the contextual relationships among all access constraints and actions via operators such as Boolean logic. To the best of our knowledge, we are the first to consider the access procedure control in order to guarantee access compliance throughout the whole process of access activities. 


Third, TBAC makes use of transferable tokoins to represent access powers. This permits high user flexibility while preserving strong security by first adopting a blockchain for secure tokoin storage and atomic tokoin transfer and then employing a TEE-enabled trusted access control object to verify whether or not the access conditions and  procedure follow the predefined \emph{4W1H} fine-grained access policy. No matter who possesses the tokoin, only the desired subject under the user-defined access conditions can redeem access. In summary, TBAC achieves fine-graininess, strong accountability, and access procedure control by adopting tokoins that define the \emph{4W1H} access policy, and the blockchain and TEE technologies to securely log all on-chain/off-chain activities. TBAC is secure against various attacks such as tokoin forgery, theft, and access policy violation.

Fourth, a complete and effective access control scheme should implement the three processes of Authentication, Authorization and Auditing to ensure that an authentic subject is authorized to access the right amount of resource, no more no less, under verifiable conditions following a fully accountable procedure. We detail all the tokoin function implementations to demonstrate how these three processes are realized and fully protected in TBAC. We also provide Go-Tokoin and Ethereum-Tokoin, the two prototypes of the TBAC scheme, with the former following the native design (Tendermint-BFT) for best performance and the latter showing the adaptivity of TBAC to mainstream blockchain platforms (Ethereum). The TEE-enabled access control object is implemented in ARM Cortex-M33 based microcontroller protected by the ARMv8-M TrustZone, to securely sample the physical environment for access condition verification and access procedure monitoring. To the best of our knowledge, we are the first to develop on the Cortex-M series TEE microcontrollers, which are specifically designed for low-cost trustworthy embedded systems by supporting hardware-level program security isolation. As the Cortex-M series TEE microcontrollers offer very little usable libraries, we build our access control object roughly from the bare metal level, thus facing a great engineering challenge. A friendly TBAC Android App is also developed for convenience and better user experience. All codes are open-sourced and available at Github.


Lastly, we present a TBAC-assisted in-home cargo delivery case study to demonstrate how TBAC is utilized for real world applications. This case study permits autonomous in-home delivery, in which a deliveryman can open the smart door and put the cargo down on its own in the designated area (e.g., mud area), and the access procedure control guarantees that the deliveryman's actual behavior does not violate the access policy (e.g., not entering the main room) and ultimately the home owner's physical security. More importantly, this case study provides a seamless integration of blockchain and IoT to extend the on-chain trustworthiness to off-chain physical world. This piece of work itself has its own significance in many IoT applications that require trusted management. Our case study also shows that secure and accountable accesses to physical resources can be achieved by techniques that are used to be only available in the digital world. 

The paper is organized as follows. We first present the most related work and analyze the existing challenges in Section~\ref{sec:Related}. Then we propose our tokoin based access control model TBAC in Section~\ref{sec:TBAC} and detail its implementation in Section~\ref{sec:implementation}. The TBAC-assisted in-home cargo delivery case study is presented in Section~\ref{sec:casestudy}. We  conclude this paper with a future research discussion in Section~\ref{sec:conclusion}.

\section{Background and Related Work}
\label{sec:Related}

In this section, we first introduce two major categories of existing access control schemes, namely Access Control Matrix (ACM) and Capability-based Access Control (CapBAC); then we discuss their design deficiencies, analyze how these deficiencies jointly cause the overprivilege challenge and how the actual Industrial implementations worsen the corresponding problem; finally we present the current countermeasures to tackle the challenge. 



\subsection{Mainstream Access Control Schemes}
There exist many access control schemes such as Role-based Access Control (RBAC), Attribute-based Access Control (ABAC), Access Control Matrix (ACM), Access Control Lists (ACL), and Capability-based Access Control (CapBAC). Most of them can be classified into two categories, namely Access Control Matrix (ACM) and Capability-based Access Control (CapBAC)\cite{indu2018identity} \cite{harrington2003cryptographic}.

\subsubsection{Access Control Matrix (ACM)}
An ACM refers to a table that lists precise security entries recording which user or group has which privileges over what resources. These security entries are also called security references. A reference server stores these references and upon access request, it looks up the references and judges whether the access requester passes the reference check or not. Typical examples of ACM include the Access Control List (ACL) from the Unix file system, which defines whether or not a specific user has Read/Write/Execute privileges over a file, and the Role-based Access Control (RBAC), which extends the minimum subject unit from an individual to a user group. RBAC is proved to be equivalent to ACM with respect to the policies they can represent \cite{saunders2001role}. Attribute-based Access Control (ABAC) is an improvement over RBAC that requires more contextual attributes from the access subject, and combines them with boolean logic to achieve a finer granularity of access control\cite{hu2013guide}. As ABAC explicitly stores policy scripts in an access matrix \cite{karjoth2008implementing}\cite{jahid2009enhancing}, we classify it to the ACM category.

\subsubsection{Capability-based Access Control (CapBAC)}

CapBAC makes use of a certificate-like capability (token) to represent the access privilege and realize direct peer-to-peer access control. After authentication, the access server issues a token, possessing which gives right to access certain resources within a certain period of time. This is more widely used in Industry, and the most popular CapBAC schemes include OAuth, OAuth2, and JSON Web Token (JWT). 
There also exist mechanisms that decentralize CapBAC, which are mainly based on the simple marriage of blockchain technologies and CapBAC schemes \cite{xu2018blendcac}.

\subsection{Open Challenges} \label{sec:existing:challenges}

\subsubsection{Coarse Granularity}
Both ACM and CapBAC have pros and cons. ACM is featured by precise definitions of simple access privileges such as read or write, and is easy to implement for a small scale system. However, as the number of users and the granularity of access conditions increase, higher-dimensional data is needed, making ACM excessively larger and sparser. To keep ACM practically small, the operations and the policy granularity must be limited. Obviously, this makes it hard to release a precise amount of resource, hence causing the rise of the \textbf{coarse granularity problem} where the access target and method are defined vaguely. This coarse granularity is attributed to ACM's user-centric nature, and its inability to depict complex contextual relationships. An exemplary attempt to overcome the coarse granularity challenge is ABAC, as it adds more attributes and permits contextual relationships among the attributes through boolean logic combinations. Note that ABAC is close to a satisfactory solution against coarse granularity, but it suffers from high complexity and verbosity, making it increasingly dim for future worldwide adoptions \cite{XACMLisdead}\cite{lorch2003first}. CapBAC provides an unforgeable certificate-like capability (token) to prove an entitled access power, but it also puts up with the coarse granularity problem. 
This is because the creation of access tokens does not mandate a fine-grained access policy. For this reason an OAuth token issued to access \textit{one photo} in a smartphone can be maliciously (and easily) exploited to access \textit{all photos} in the smartphone \cite{fernandes2018decentralized}.

\subsubsection{Weak Auditability}

The weak auditability problem refers to an inability to monitor all access control activities, including authorization, authentication, redemption, etc. In ACM, most access auditing logs are recorded on the victim device, thus an attacker can use the excessive control power obtained from unauthorized attacks to easily (and in fact, commonly, found by Cozzi \emph{et al.} \cite{cozzi2018understanding}) delete the logged evidence without being caught. To make things even worse, a large portion of devices and platforms such as Nest and Samsung SmartThings in practice, even do not log access activities  \cite{schuster2018situational}.

CapBAC is also vulnerable to the weak auditability problem. In most mainstream projects and standards, once an access capability is issued, no constraint is placed to protect the use of such capability -- the access privilege is technically unlimited. We take the most popular and widely used CapBAC protocol OAuth 2.0 as an example\footnote{OAuth 1.0 is proved to be insecure and deprecated \cite{hardt2012oauth}}: after initial authorization, the OAuth protocol creates a non-cryptographic token and transfers it through SSL/TLS to the subject, requiring the subject to be responsible for keeping the token confidential. Nevertheless, in all CapBAC schemes, whoever possesses the token is assumed to be the legit access subject, and can access the resource without further authentication and auditing. As a result, it is easy to impersonate or steal a token to perform unlimited amount of unauthorized accesses without being detected \cite{hu2014application} \cite{fernandes2016security}. 


\subsubsection{Lack of Access Procedure Control}
The lack of access procedure control is a systematic neglect in both ACM and CapBAC. This problem is caused by the current access decision-making process that requires only an instantaneous check on access conditions at the moment when the access is requested. After a decision is made, the task of access control is accomplished, leaving a lot of room for the actual access behaviors to completely go wild. Therefore it is not surprising to observe that  most, if not all, overprivilege attacks are explicitly discoverable during the access process. The lack of  access procedure control may not seem to be a big problem in classic computer systems in which an access control scheme mainly plays the role of a door lock safeguarding the jewels inside a cabinet, assuming that a subject is the legal owner of all jewels inside the cabinet as long as the subject provides the right code to  open the lock.  But for IoT applications that are deployed in open and distributed environments in recent years, where an access event typically involves a procedure of manipulating different resources with different access rights at different instant of time, access condition verification alone is insufficient and an effective access procedure control is obviously needed.


\subsubsection{Overprivilege Challenge:}
Obviously, when an access control scheme suffers from coarse granularity, it is easy to release more resource than one needs. This directly gives rise to the so-called overprivilege challenge, which can be further worsened by the weak auditability problem, where an access scheme fails to audit all activities,  and the lack of access procedure control, which gives unlimited opportunities for the attacker to manipulate. Overprivilege has been a popular open research topic for a long time \cite{tian2017smartauth}\cite{fernandes2016flowfence}\cite{sikder20176thsense}, yet there still lack effective solutions. 

With the fast and prevalent penetration of IoT applications in our daily life, the overprivilege challenge becomes increasingly grave and intolerable, especially when considering the situations where relevant cyber threats are extended onto our real-life security. Imagine how threatening it can be in a world where Alexa’s red light goes on to secretly record our daily conversations and the smart door lock is opened up by an adversary behind the scenes. Felt \textit{et al.} found that overprivilege vulnerabilities exist in over one-third of the Android-driven IoT devices \cite{felt2011android}, and Fernandes \textit{et al.} reported that in Samsung SmartThings platform over 55\% of SmartApps are overprivileged due to coarse access control granularity \cite{fernandes2016security}. Zhang \emph{et al.} discovered that attackers can create malicious apps on the Amazon skill platform to perform overprivilege attacks and conduct eavesdropping activities through Alexa \cite{Zhang2019dangerous}. Such attacks have occurred repeatedly on many off-the-shelf products such as Amazon Alexa and Google Home all over the United States, and have been reported by many mainstream public media \cite{usatodayalexa} \cite{washingtonpostalexa} \cite{cnngoogle}. 

In summary, due to the lack of an accountable and fine-grained access control scheme that can also monitor the access procedure, an adversary can easily over-access the designated resources, or even perform unauthorized accesses. One can see that overprivileged accesses wreak havoc, thus effective solutions are desperately needed.


\subsection{In Practice: Industrial Implementations}

In Industry, the common practice of IoT access control is either \emph{open-port direct control }or \emph{server-assisted remote control}. These two common implementations largely ignore the three requirements of fine-graininess, strong auditibility, and access procedure control.

\begin{figure}[!htb]
	\centering
	\includegraphics[width=0.48\textwidth]{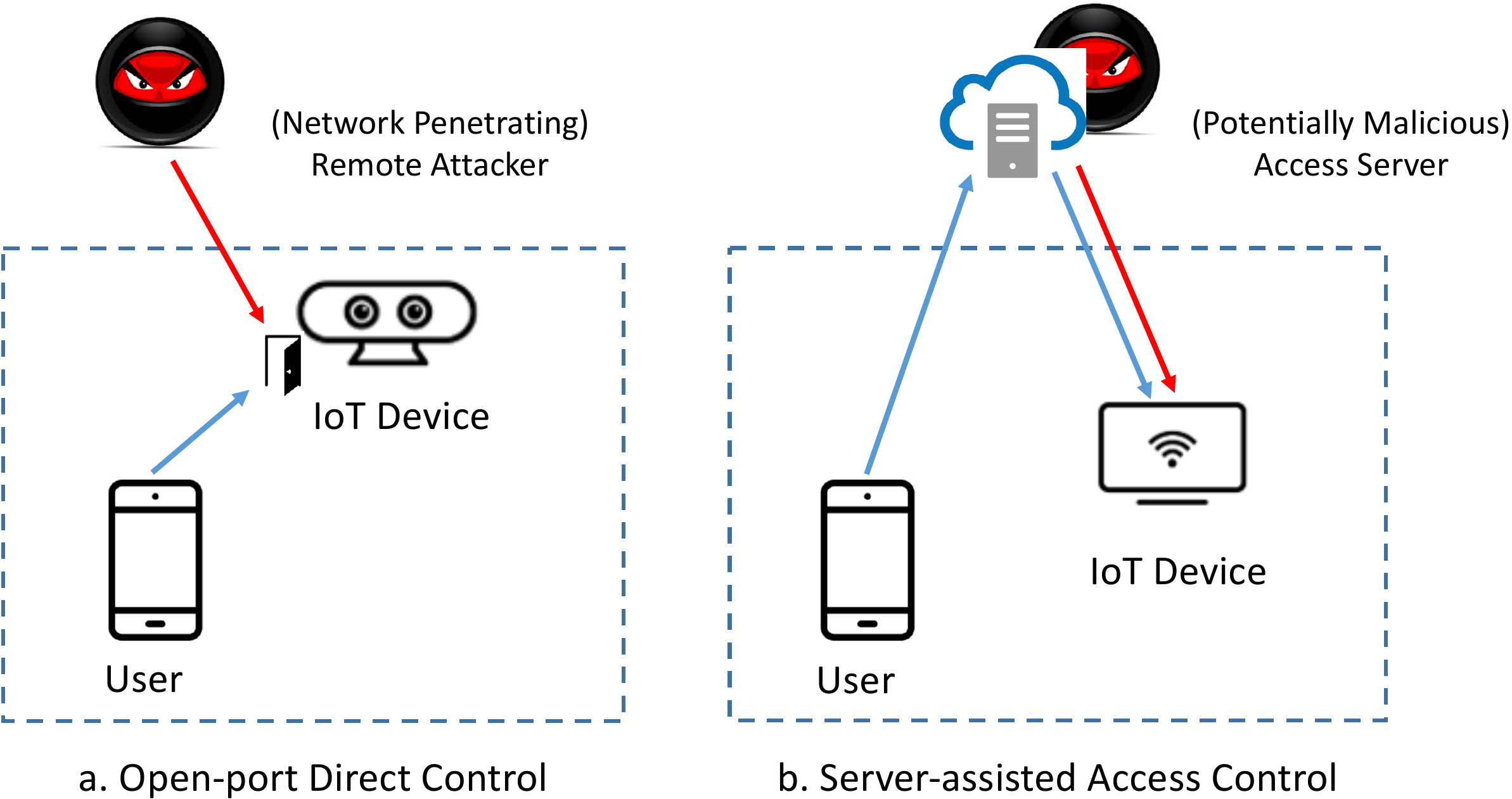}
	\caption{Two Major Types of Industrial Implementations on IoT Access Control}
	\label{fig:twoaccess}
\end{figure}

\subsubsection{Open-port Direct Control}
As shown in Fig. \ref{fig:twoaccess}a, the open-port direct control requires IoT devices and their drivers' network ports open at all times; thus any user can freely connect and control the devices without authentication. It was originally proposed to decrease authentication latency and implementation complexity, and had once become increasingly popular until the first IoT botnet Mirai exploited such vulnerability and compromised more than 65,000 IoT devices within the first 20 hours after its initial release in August 2016, with over 600,000 devices infected at peak \cite{antonakakis2017understanding}. Note that Mirai simply flooded SYN messages on TCP port 23 (telnet) and achieved such results. Even now more than 62\% of IoT devices tested were still found vulnerable to remote access attacks exploiting open ports and weak credentials 
\cite{F5Labs}. It is clear that this open-port direct control implementation is near-defenseless, not to mention the close-to-none auditability, fine-granularity, and access procedure control. Therefore many mainstream manufacturers switched to the server-assisted remote control in recent years.

\subsubsection{Server-Assisted Access Control}

As shown in Fig. \ref{fig:twoaccess}b. IoT devices are connected to their access servers that are run by their manufacturers/service providers. If a user wants to access and operate its device, a request needs to be sent to the access server. Then the server authenticates the user’s identity with typically a username-password pair, and performs controls towards the IoT device \textit{on behalf of} the user. From a technical perspective, the access control decision is made by the server, not the user, and the actual operating commands can only be issued from the access server. 

It may sound weird that the owner of a device does not have actual access power over the device, but it’s true. Most off-the-shelf IoT devices such as Samsung SmartThings, Apple Homekit or Amazon Alexa, use a close-source platform to support very limited programming towards the devices, thus greatly limits the access policy granularity \cite{fernandes2016security}. On the other hand, with the access server having the technically highest control privilege, the service providers can only make legal-level promises against unauthorized accesses \cite{amazoninhome}. If the access server itself is compromised by an adversary who then commits an unauthorized access activity, it can erase the log evidence with root privileges \cite{cozzi2018understanding}; therefore auditability is very vulnerable in server-assisted access control. Additionally, as in open-port direct control, access procedure control cannot be realized in server-assisted remote control.

\subsection{Research Countermeasures}

\subsubsection{Security Analysis-Based Countermeasures}
Most countermeasures that attempt to mitigate the overprivilege challenge are based on security analysis, in which the overprivilege vulnerabilities are inspected as programming deficiencies and  identified based on the learned characteristics. Fernandes \emph{et al.} uncovered a severe overprivilege vulnerability in the Samsung SmartThings platform, which allows attackers to falsely turn on a fire alarm \cite{fernandes2016security}, but they did not propose any effective defense mechanism. Jia \emph{et al.} presented a graph-based algorithm to automatically excavate the overprivilege weaknesses in a smart home system \cite{jia2018novel}. Celik \emph{et al.} hand-crafted 20 common flawed apps, with which a model-checking based solution was developed to automatically identify the flaws \cite{celik2019iotguard}. In summary, all the mechanisms mentioned above mitigate the overprivilege challenges by first discovering the vulnerabilities, then learning features from them, and finally detecting their presence in a general environment. Their false-negative rates are usually high, and emerging vulnerabilities (e.g., zero-day vulnerabilities) are hard to be detected.

\subsubsection{Advancement in Access Control}
Blockchain can provide a trustworthy environment for many applications, ranging from secure transactions \cite{nakamoto2019bitcoin} to trusted verifiable computing \cite{kumaresan2014use}. Apart from the access control works mentioned above, researchers also resorted to blockchain technologies. However, current blockchain-based access control schemes are mostly the simple marriages of blockchain and existing solutions, thus suffering from the same open challenges. 

For examples, Pinjala \textit{et al.} proposed a decentralized framework based on IOTA blockchain to create and manage the access capability tokens \cite{pinjala2019dcaci}; Zhang \emph{et al.} developed an ACL-based access control scheme on top of multiple Ethereum smart contracts \cite{zhang2018smart}, following the classical coarse-grained Unix style $\{\textit{Read, Write, Execute}\} \rightarrow \{\textit{Allow, Deny}\}$; Xu \emph{et al.} presented BlendCAC, which employs Ethereum smart contracts in replacement of the traditional capability access server to issue and manage coarse-grained Linux-style access capabilities \cite{xu2018blendcac}; 
Maesa \emph{et al.} implemented a blockchain-based ABAC scheme, where a blockchain is used to record the location of an externally stored ABAC policy \cite{maesa2018blockchain}. 
Although the scheme in \cite{maesa2018blockchain} has a better policy granularity, they all suffer from other problems of overprivilege challenge, weak auditability, and lack of access procedure control.

The status-quo urges us to ponder the following question: can we design an accountable and fine-grained access control scheme that can precisely define the access privilege and control the access procedure while enforcing high auditability on all operations towards an access activity? 
To answer this question, we strive to design a secure, accountable and fine-grained access control scheme that allows owners to have cryptographic level security confidence over the uses of their resources without crossing the access policy boundaries defined by themselves. We leverage blockchain and cryptographic primitives to instantiate access capabilities into secure and accountable cryptographic assets, namely tokoins, and design TBAC, a fine-grained and accountable access control model that supports access procedure control, to overcome all the challenges analyzed above and realize the design goals of effective access control.  


\section{TBAC: Tokoin-Based Access Control}
\label{sec:TBAC}

Tokoin-Based Access Control (TBAC) is a general access control model that employs cryptographic coins to represent access privileges. Tokoin is termed by combining ``token'' and ``coin'', for the purpose of materializing access capabilities as atomic, accountable, and transferable digital assets. In this section, we present our design goals, security assumptions, threat model, and an overview on TBAC.

\subsection{Design Objectives, System Assumptions, and Threat Model}
\label{sec:objective:assumptions:threat}

\subsubsection{Design Objectives}
The major objective of TBAC is to provide a \emph{flexible} and general access control model that can offer \emph{fine-grained} and \emph{fully-accountable} control over the access of private digital or physical resources. TBAC should be \textit{flexible} enough to support simple access control tasks that mainly require identity authentication, as well as complex ones that require external environment data inputs. It should be able to precisely define \emph{who}, \emph{what}, \emph{where}, \emph{when}, and \emph{how}, the five critical elements that determine the \emph{fine-graininess} level of an access control scheme, to answer the question of \emph{who is allowed to do what in where at when by how}. This is termed as the \emph{4W1H} access policy, where the four \emph{W}'s specify the access conditions that must be met in order for an access request to be granted while the \emph{H} describes the exact access procedure that must be strictly followed during access. Only the resource owner has the right to define and modify a \emph{4W1H} access policy, and its five elements can be dynamically changed such that the owner can flexibly control the access to its resource at its own will. In TBAC, this \emph{4W1H} access policy is minted into a tokoin, and within its whole lifetime, all actions towards the tokoin are securely logged for \emph{auditing} purpose.

TBAC should comply with the AAA requirements, i.e., it must clearly and precisely define the Authentication, Authorization, and Auditing processes that should be implemented by all complete and secure access control schemes from the information security perspective \cite{sandhu1996authentication}. The Authentication process is an act of establishing or confirming the identity or capability as authentic; the Authorization process determines whether a person or a process is authorized to perform a given access activity; and the Auditing process logs all access events that have security significance. 

\subsubsection{High-Level System Requirements}
\label{sec:assumptions}

To achieve the objectives mentioned above, TBAC makes two important system requirements to assure security. 

First, TBAC relies on a database that supports secure tokoin storage, atomic tokoin transfer, and accountable activity auditing. Such properties are very similar to what blockchain can provide. As a result, blockchain is adopted in TBAC. More specifically, the existence of a blockchain system that is provably secure for trusted storage (fork-free) and state transitions is assumed by TBAC. It is employed to manage all operations over a tokoin as transactions such that the tokoin can be securely stored and all (on-chain) activities over it can be logged for auditing purpose.

Second, TBAC requires a reliable access control object that works on behalf of the resource owner to verify tokoin capability validity, check if the fine-grained access conditions are met, make actual access decisions, and monitor the access procedure. This requires the access control object to have secure connections with the blockchain for tokoin reception and with the resource for access control. Considering the AAA requirements, RACO needs to establish a trust environment that directly extends that of the blockchain, i.e., extending trust from on-chain to off-chain, such that one can be sure that the access policy carried by a tokoin is strictly followed during access and all access-related activities are securely logged. To achieve this objective, we adopt a Trusted Execution Environment (TEE), a special kind of trusted hardware that has a tamper-proof area for secure programs, to host the access control object and other secure software in TBAC for information collection to realize secure and accountable access control. 

Conceptually, TEE is a tamper-resistant processing environment enabling isolation and secure storage within a chipset \cite{sabt2015trusted}. Technically speaking, a TEE implements fundamental functions such as secure boot, runtime isolation, secure storage, secure scheduling, trusted I/O, and trusted remote management.  
The program executed in TEE resists software attacks as well as physical tampering such that the authenticity, integrity, and confidentiality of the codes and runtime states are guaranteed. In this paper, we assume that long-range vulnerabilities do not tamper programs within the TEE secure zone, and that any local program can be executed as expected. However, physical damages and attacks towards TEE hardware are out of the scope of this paper, thus will not be considered.

\subsubsection{Threat Model}\label{sec:threat:model}

In this paper, we consider a general but powerful adversary $\mathcal{A}$ who knows the details of the TBAC construction procedure, has access to all public functions, and can read public requests and messages. $\mathcal{A}$ may compromise a fraction of the participants but this fraction is not large enough to thwart the normal secure operations of the underlying blockchain system. 
It aims to perform an overprivileged access (an unauthorized access is deemed as a special kind of overpriviledged access) to the private resource. In practice, this adversary can be a remote attacker that can penetrate the local network, or a local attacker, or the hosting access server of the resource, which are popular in current IoT industry.

\begin{figure}[!ht]
	\centering
	\includegraphics[width=0.45\textwidth]{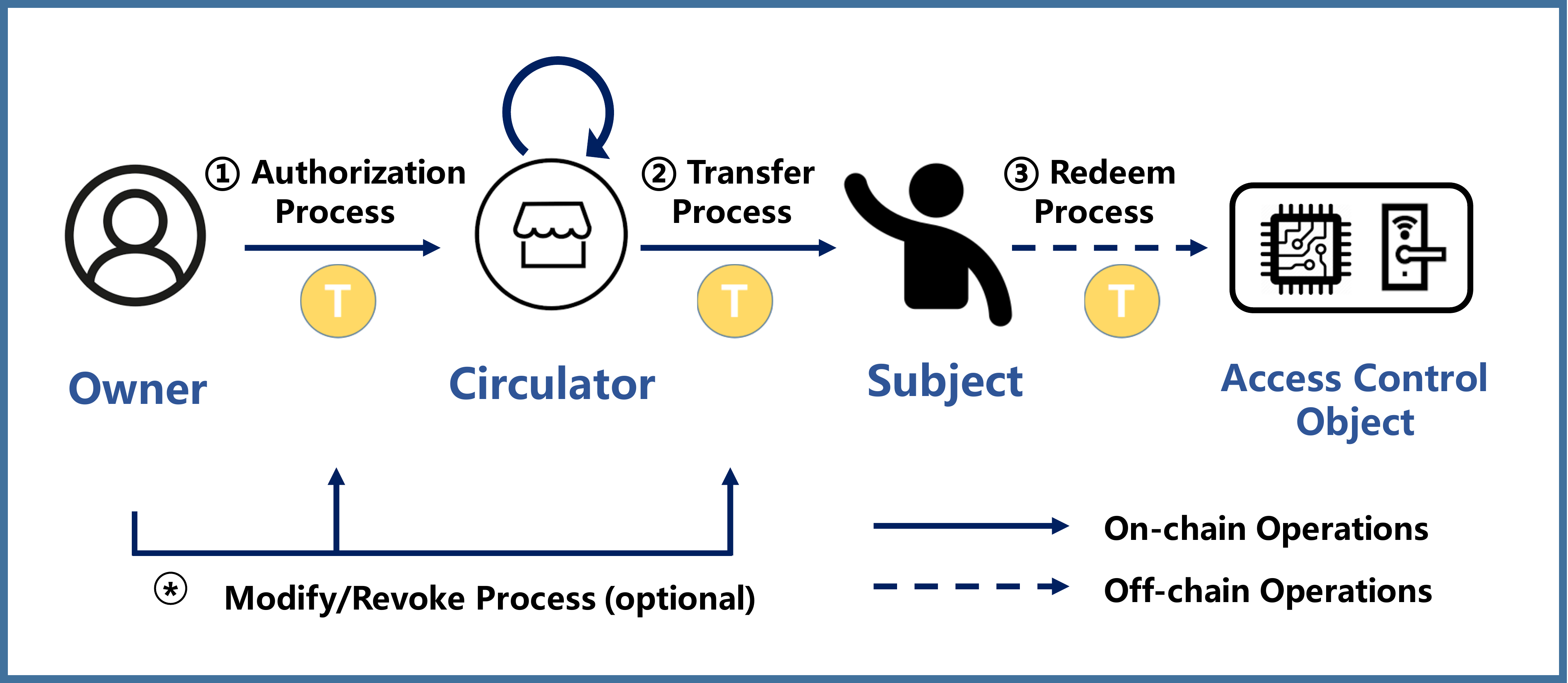}
	\caption{Overview on TBAC}
	\label{fig:overview}
\end{figure}

\subsection{An Overview on TBAC}

\subsubsection{Entity Roles in TBAC}
\label{sec:roles}
Typically an access control scheme consists of the following entity roles. An \textit{owner} owns a piece of private \textit{resource} $D$, which could be a hardware device or a software program. The owner authorizes a \textit{subject} to access its resource under a series of restrictions including access conditions and an access procedure, which together define the \textit{access policy}. In TBAC we have a more refined definition of the entity roles, denoted by $\mathcal{R} = \{ {r_O, r_S, r_C, r_{\textit{ACO}}} \}$, where $r_O$ is the resource owner who issues a tokoin $t$ along with an access policy $t.\textit{policy}$ for its resource $D$, $r_C$ is the circulator, who must be the current holder, of the tokoin $t$ that helps transfer $t$ to a legit subject $r_S$. An access control object $r_{\textit{ACO}}$ is the local robust access control module that actually controls the access to $D$ on behalf of $r_O$. 
Upon subject $r_S$'s tokoin redemption request, $r_{\textit{ACO}}$ verifies if the tokoin capability is valid and if the conditions required by the access policy are met; and if yes, grants the access to $r_S$ and monitors the access procedure. Note that $r_C$, the circulator and current holder of $t$, could be the $r_O$, or a $r_S$, or any other legitimate participant in TBAC. Also note that $r_S$ does not have to be one specific participant -- it could be a group of participants possessing certain legitimacies to access the resource $D$. We further emphasize that the minimum atomic access privilege in TBAC is a tokoin capability, which is an actual instantiation of the privilege. Within a tokoin access activity, a participant can only have one role. However, it can participate in multiple activities of different tokoins, thus can have a different role in each respective activity at any instant of time. In TBAC, a participant can be uniquely identified by its address or a public key issued by the system. 


\subsubsection{TBAC Description}

TBAC intends to let the resource owner completely control the access privileges over its resources. To achieve this goal, it manages the access control powers in the form of secure, atomic, accountable and policy-programmable cryptographic coins. This approach is featured by four properties: only the resource owner can create, modify, or revoke a tokoin in order to completely control the access to its resource; the current holder of a tokoin can atomically transfer the tokoin but only a legit subject can redeem the tokoin; all conditions required by the access policy must be met upon redemption and the exact access procedure defined by the access policy must be strictly followed for correct resource access; all access activities must be audited in a secure, accountable, and fine-grained way.

Fig.~\ref{fig:overview} illustrates the operations of our TBAC model. 
When the owner $r_O$ of resource $D$ wants to authorize a subject $r_S$ to access $D$, $r_O$ drafts an access policy and securely creates a tokoin. Then the owner $r_O$ transfers the tokoin to the first circulator $r_C$, as demonstrated by the Authorization Process \circled{1} in Fig.~\ref{fig:overview}.  

The tokoin circulator $r_C$, if it is not the final subject, accountably transfers the tokoin to another participant when needed. This can help the resource owner to deliver the access capability to a legit subject, or dynamically delegate a different group of subjects to access its resource during the tokoin transfer process as the resource owner can modify a tokoin before it is redeemed (labeled by a \circled{$*$} in Fig.~\ref{fig:overview}). 
However, although the transfer of a tokoin is allowed between any two participants, the final tokoin redemption must be performed by a legit subject. By this way one can permit a great access privilege management flexibility while still guaranteeing that the final access activity strictly follows the owner's discretion. This can motivate new methods in physical resource management such as allowing secure resource rent, trading of hardware/software resource right-of-use, or assigning someone to run an errand more flexibly. We provide a case study where autonomous secure in-home parcel deliveries are enabled for the convenience of ordinary people. This is the Transfer Process \circled{2} in Fig.~\ref{fig:overview}. 

Finally, when a legit subject $r_S$ possesses the tokoin and wishes to redeem for resources, the redemption process checks if all conditions defined in the access policy are met to ensure that it is indeed a legit subject to access the right resource under correct conditions. The redemption process also needs to monitor the access procedure, if defined by the access policy, to make sure that the procedure is strictly followed. Such a design can guarantee that despite the flexibility of the access power transfer process, the final redemption must follow user-defined access rules. Any violation per the access policy renders the access fail. 
This is the Redemption Process \circled{3} in Fig.~\ref{fig:overview}.

We would like to emphasize that an access policy carried by a tokoin can be flexibly defined at any graininess level per the resource owner's preferences/needs, by inputting values at different granularity to the  \emph{who}, \emph{what}, \emph{where}, \emph{when}, and \emph{how} fields. Moreover, the \emph{who} field in the access policy can specify a group of legit subjects to access the resource as the resource owner may not be aware of the exact subject when the tokoin is minted. Additionally, the owner can modify any component of the access policy in a tokoin or simply revoke the tokoin as long as it is not redeemed, providing strong fine-grained control over its resource access privileges at any instant of time. On the other hand, one can see that TBAC is applicable to the access controls over both digital resources as well as physical resources, as the five fields of a \emph{4W1H} access policy do not rely on any resource-specific feature. Nevertheless, as a general fine-grained access control model, TBAC needs to be instantiated differently for different applications. We detail the TBAC implementation in the next section and present a case study to demonstrate its applications in Section \ref{sec:casestudy}.  




\subsubsection{The Properties of Tokoin in TBAC}


To achieve the goal of providing fine-grained accountable access control, a tokoin in TBAC must possess the following properties:
%
\begin{description}
    \item[Security]: A tokoin is a materialized entity of a secure and accountable access power, in the form of a cryptographic coin. It cannot be falsely created, tampered with, used, or revoked by an adversary.
    \item[Atomicity]: The transfer of a tokoin requires the simultaneous arrival at the receiver and removal from the sender, making it hard to double-redeem the tokoin. 
    \item[Strong Auditability]: All current or historical activities over or changes to a tokoin, as well as all activities during the access to the resource based on the policy carried by a tokoin, are securely audited. 
    \item[Fine-Grained Access Policy]: The access policy carried by a tokoin specifies \emph{4W1H} 
    such that a user can precisely describe both the access conditions (the \emph{4W}) to be checked before granting the access, and the access behavioral constraints (the \emph{1H}) after granting the access.
    The fine granularity level is exclusively determined by the resource owner.
    \item[Access Procedure Control]: A tokoin should carry access conditions for pre-access compliance verification and access restrictions for during-access behavior obedience monitoring.    

    
    \item[Direct Access Privilege Sovereignty]: The access privilege over a piece of resource can only be defined and issued by the resource owner via a tokoin at its discretion when needed, without intervention of any third party access server.

    \item[Transferability of Access Power]: A tokoin can be transferred peer-to-peer from one holder to another through a properly audited procedure, if needed. 

    \item[Dynamic Access]: The access privilege carried by a tokoin can be modified or revoked at anytime and these actions should be performed only by the tokoin owner. 
    
    \item[Definite-amount of Access]: Obviously, the number of times a tokoin is allowed to be redeemed can be easily specified by the access policy, successfully preventing the tokoin from being unlimitedly used. 
    
\end{description}

One can see that there exist five operations over a tokoin: $\mathsf{Create}$, $\mathsf{Transfer}$, $\mathsf{Modify}$, $\mathsf{Revoke}$, and $\mathsf{Redeem}$. When a tokoin $t$ needs to be created, the resource owner sends an authenticated $\mathsf{Create}$ message carrying the access policy to the blockchain, which verifies the message, creates a tokoin and assembles it as a transaction, then places it into the next block. When $t$ needs to be transferred, modified, revoked, or redeemed, a corresponding authenticated message needs to be sent to the blockchain, commanding the blockchain to transfer, modify, revoke or redeem $t$. More specifically, all the tokoin manipulation functions are performed on-chain by passing authenticated messages, and each operation is recorded as a transaction such that all operations over a tokoin can be securely logged for auditing purpose. 

Note that the AAA processes are realized by the five tokoin manipulation functions with the assistance of the underlying blockchain and reliable access control object via authenticated messages in TBAC. In the following section we present a detailed implementation of these tokoin manipulation functions to illustrate how AAA are achieved.

\subsubsection{Security Discussions}

Intuitively, TBAC may suffer from two attacks: tokoin forgery and access policy violation. As mentioned in Section~\ref{sec:assumptions}, TBAC makes two system requirements: 1) a blockchain distributed ledger that is provably secure to support trusted storage and atomic state transitions; and 2) a reliable access control object $r_{\textit{ACO}}$ that can securely retrieve access policies from a tokoin in blockchain, make access decisions according to the access conditions, and monitor the access procedures. As all operations of a tokoin are performed on-chain and all messages are authenticated via provably secure signatures, the first requirement guarantees that an adversary cannot forge a tokoin by spoofing a legitimate signature to create a tokoin for or transfer a tokoin to itself, or fork a chain to include its own tokoin making the forked chain a legitimate one. On the other hand, the second requirement ensures that the access policy defined by the resource owner is strictly followed, no more no less. Based on these considerations, one can safely conclude that TBAC is secure against tokoin forgery attacks and access policy violation attacks.

\section{TBAC System Implementation}
\label{sec:implementation}

In this section, we detail an implementation of the TBAC model. We first provide a system description on TBAC, then present its Authentication, Authorization, and Auditing processes, and finally describe a native and a more adaptive TBAC prototype implementations, followed by an Android App for regular users. 


\subsection{TBAC System Description}

TBAC employs a blockchain to fulfill the security requirements of secure tokoin storage, atomic tokoin transfer, and accountable activity auditing, and a trusted execution environment (TEE) to collect trusted environment evidence and make correct final access control decisions.

\subsubsection{Primitive Elements and Functions}
\label{sec:TBAC:definition}

A tokoin $t=(t_{\textit{ID}}, pk_O, pk_H, \textit{policy}, \textit{isValid})$, where $pk_O$ and $pk_H$ are respectively the public keys of the owner and current holder of $t$, $t_{\textit{ID}}$ is a number uniquely identifying $t$ among all the tokoins generated by the owner $pk_O$, \textit{policy} defines \emph{who is allowed to do what by how in where at when}, and $\textit{isValid}$ is a binary indicator with $\textit{isValid}=1$ if and only if $t$ is still valid (not redeemed and not revoked). Note that a tokoin $t$ is uniquely identified in TBAC by the two-tuple $pk_O$ and $t_{\textit{ID}}$, denoted by $pk_O||t_{\textit{ID}}$, as two tokoins generated by different participants may have the same $t_{\textit{ID}}$. 
For conciseness, $\textit{isValid}$ is omitted and $t$ is used instead of $pk_O||t_{\textit{ID}}$ to identify a tokoin, if clear from context.

A fine-grained access $\textit{policy}$ answers \emph{who is allowed to do what at when in where by how}, 
where \emph{who} is a legit subject that may not be known before tokoin $t$ is minted, thus we employ a cryptographic $\textit{Accumulator}$ in $t.\textit{policy}$ to define a group of subjects who are allowed to redeem $t$ and access the resource \emph{what}. The spatio-temporal access constraints, i.e., \emph{when} and \emph{where}, can be sampled from a device such as a GPS receiver at redemption. The access procedure is strictly defined by \emph{how},  
which can be specified by \emph{atomic actions} to manipulate the resource. Whether or not such a procedure is strictly followed can be monitored by a software program or a hardware device such as a video camera. Note that one can have multiple atomic actions and constraints as well as their relationships (e.g., ``and'', ``or'', ``not'') defined in $\textit{policy}$, capturing the contextual relationships,  to precisely describe under which condition and how the resource should be accessed, to realize access control at the user-defined graininess level.  
%
%

TBAC relies on a blockchain public ledger for secure storage and atomic state transfer. The ledger is maintained by all the blockchain nodes participating in consensus and block construction. A tokoin is stored on-chain and all the tokoin manipulation operations are performed on-chain via function calls that are realized by passing messages, i.e., the function caller sends a message to all blockchain nodes for tokoin $\mathsf{Create}$, $\mathsf{Transfer}$, $\mathsf{Modify}$, $\mathsf{Revoke}$, and $\mathsf{Redeem}$. All the messages must be signed by the function caller's secret key and all blockchain nodes receiving the message must first verify its authenticity based on the carried signature. In TBAC, a message from caller $pk_i$ has the following format:
\begin{equation}
    msg:[t, op, \{\textit{policy}\}, \{pk^\prime\}]_{\sigma_{pk_i}}
\end{equation}
where $op$ is the operation code that distinguishes which function to be called for tokoin $t$, the braces contain optional information, with $\{\textit{policy}\}$ holding a valid policy if $op=\textit{create}\ \text{or}\ \textit{modify}$ and $\{pk^\prime\}$ being a new receiver's address if $op=\textit{transfer}$, and $\sigma_{pk_i}$ is the message signature signed by the \emph{secret key} of the function caller $pk_i$ for message authenticity verification. Note that the message actually includes the composite ID $pk_i||t_{\textit{ID}}$ of $t$ but we use $t$ instead of  $pk_i||t_{\textit{ID}}$ here for better clarity.  

%
%
%

\begin{definition}
\normalfont
    For a set of participants $\P=\{P_1,\ldots,P_n\}$, TBAC defines
	the following nine functions:
	
	\textbf{Gen:} The function $\mathsf{Gen}$ generates keys for the participants in TBAC. When called by participant $P_i\in\P$, $\mathsf{Gen}$ produces a public/secret key pair $(pk_i,sk_i)$, returns $sk_i$ to the caller and broadcasts the public key ${pk}_i$ to all $P_j \in \P$. Note that $pk_i$ can be used to uniquely identify $P_i$.
	
	\textbf{Verify:} The function $\mathsf{Verify}$ supports overloading. On inputs of $\sigma$ and $pk$, it outputs 1 if and only if signature $\sigma$ was indeed signed by participant $pk$; on input of $t$, it outputs 1 if and only if tokoin $t$ is still valid ($t.\textit{isValid}=1$); and on inputs of $t$, $pk$, and $\textit{role}$, it outputs 1 if and only if participant $pk$ holds the $\textit{role}$ in tokoin $t$. 

	\textbf{Create:} The function $\mathsf{Create}$ creates a new tokoin $t$. 
	On input of an access \textit{policy}, it outputs a tokoin $t=(t_{\textit{ID}}, pk_{i},$ $ pk_i, \textit{policy}, \textit{isValid=}1)$ to the caller $P_i$, who then acquires role $r_O$. 
	
	\textbf{Transfer:} The function $\mathsf{Transfer}$ transfers a tokoin $t$ to a different holder. When called by participant $P_i\in\P$, it requires $\mathsf{Verify}(t,pk_i,r_{C}) = 1$. On inputs of a receiver $pk_j$ and a valid tokoin $t = (t_{\textit{ID}}, pk_O, pk_i, \textit{policy})$, it modifies $t$ to be $(t_{\textit{ID}}, pk_O, pk_j, \textit{policy})$, i.e., changing the current holder of $t$ from $pk_i$ to $pk_j$, 
	who then acquires role $r_{C}$. 
	
	\textbf{Modify:} The function $\mathsf{Modify}$ modifies the access policy of a tokoin $t$. When called by participant $P_i\in\P$, it requires $\mathsf{Verify}(t,pk_i,r_{O}) = 1$. On inputs of a valid tokoin $t=(t_{\textit{ID}}, pk_O, pk_H, {\textit{policy}})$ and an updated policy ${\textit{policy}}^{*}$, it modifies $t$ to be $(t_{\textit{ID}}, pk_O, pk_H, {\textit{policy}}^{*})$. 
	
	\textbf{Revoke:} The function $\mathsf{Revoke}$ revokes a tokoin $t$. When called by participant $P_i\in\P$, it requires $\mathsf{Verify}(t,pk_i,r_{O}) = 1$. On input of a valid tokoin $t$ it outputs a null tokoin $\perp$ with $\textit{isValid}=0$.
	
	\textbf{Redeem:} The function $\mathsf{Redeem}$ allows Object $r_{\textit{ACO}}$ to redeem resource $D$. When called by participant $P_i\in\P$ with tokoin $t$, it requires $\mathsf{Verify}(t,pk_i,r_{S}) = 1$. On input of a valid tokoin $t$, it releases resource $D$ and revokes $t$ if $\mathsf{PolicyCheck}(t.\textit{policy}) = 1$; otherwise, it transfers $t$ back to $r_S$.
	
	\textbf{PolicyCheck:} The function $\mathsf{PolicyCheck}$ samples the current digital or physical environment and verifies whether the redemption conditions and procedure are completely satisfied. On input of $t.\textit{policy}$, it returns 1 if and only if all the access conditions are met and the access procedure is strictly followed. 
	
	\textbf{Auditing:} The function $\mathsf{Auditing}$ logs all activities (mainly the function calls of $\mathsf{Gen}$, $\mathsf{Create}$, $\mathsf{Transfer}$, $\mathsf{Modify}$, $\mathsf{Revoke}$, and $\mathsf{Redeem}$) and writes them as a script into the corresponding transactions on the blockchain public ledger.

\end{definition}

Among these nine functions, $\mathsf{Gen}$ is used during participant registration, $\mathsf{Create}$, $\mathsf{Transfer}$, $\mathsf{Modify}$, $\mathsf{Revoke}$, and $\mathsf{Redeem}$ are used to manipulate a tokoin, $\mathsf{Auditing}$ is called by all the above six functions before their returns to log all activities for auditing purpose, $\mathsf{Verify}$ is called by the five tokoin manipulation functions to ensure authenticity and operation privilege, and $\mathsf{PolicyCheck}$ is called only by $\mathsf{Redeem}$ for redemption policy verification. 

A participate can be a regular blockchain node or a user who is able to talk with the blockchain via a secure channel. Whenever a participant needs to  $\mathsf{Create}$, $\mathsf{Transfer}$, $\mathsf{Modify}$, $\mathsf{Revoke}$, or $\mathsf{Redeem}$ a tokoin, a signed message is sent to the blockchain who then verifies the message authenticity. The capability of the function caller also needs to be examined before tokoin manipulation. 
A tokoin stays on-chain before being redeemed and a new transaction is issued whenever a function call is made for a tokoin operation. The transaction contains the tokoin as well as a script describing the calling message and the activities caused by the operation for auditing purpose.

The nine functions defined above constitute a complete TBAC scheme that complies with the AAA standard mentioned earlier to control the uses of the resource $D$. The Authentication process comprises the functions $\mathsf{Verify}$ and $\mathsf{PolicyCheck}$, with $\mathsf{Verify}$ essentially verifying the function caller's identity as well as its tokoin capability and $\mathsf{PolicyCheck}$ authenticating the current policy upon access. The Authorization process is made up of five functions $\mathsf{Create}$, $\mathsf{Transfer}$, $\mathsf{Modify}$, $\mathsf{Revoke}$, and $\mathsf{Redeem}$, with $\mathsf{Create}$ creating a tokoin defining the access privilege while $\mathsf{Transfer}$, $\mathsf{Modify}$, and $\mathsf{Revoke}$ making proper modifications to the tokoin capability, and $\mathsf{Redeem}$ finally taking back the tokoin capability and redeeming the agreed resource.  Function $\mathsf{Auditing}$ implements the Auditing process, making accountable audits over all activities and modifications to the access tokoin.


In the following subsections we outline our TBAC building blocks, detail the designs of the AAA processes, and present our prototype implementations. 

\subsubsection{Building Blocks}\label{sec:building:blocks}

In our TBAC implementation, we utilize multiple key cryptographic primitives that are previously proved secure and computationally accessible. They are carefully chosen to fulfill TBAC's security requirements while not over-qualified for the tasks and not increasing unnecessary overhead. 

\textbf{\textit{Tendermint-BFT:}}
TBAC runs on top of a blockchain system. We select a consortium (permissioned) chain setting in which the blockchain security largely relies on the consensus process (we don't consider network layer attacks). For the best collective reliability, we choose Tendermint-BFT as our consensus algorithm. Tendermint-BFT belongs to the Byzantine Fault Tolerance (BFT) consensus family, and it is an improved version over the popular Practical-BFT (PBFT). Ordinary BFTs like PBFT must know the total number of participating nodes \textit{a prior}, and may also suffer from Sybil attacks. Tendermint-BFT addresses these challenges by assigning different weights to different validator nodes (where PBFT basically assumes the same weight for every node). The weights can be quantified by stakes, or other resources or self-defined security indices. This provides both flexibility and reliability as the security anchors can be carefully selected to enhance performance. Tendermint-BFT is secure (and fork-proof) when less than one third of the nodes are malicious, according to the security analysis in Buchman \textit{et al.}~\cite{buchman2018latest}.

%

\textbf{\textit{Digital Signature:}}
Digital Signature allows anyone to verify whether a message is indeed signed by the claimed signer, and is not altered in transit. Our actual TBAC implementation employs ECDSA following the NIST standard because it uses a shorter 256-bit key and has a lighter computation burden compared to the RSA-backed DSA, thus especially suitable for embedded systems. In TBAC, signatures are adopted to verify function callers' identities for the purpose of granting them different tokoin manipulation privileges.


\textbf{\textit{Cryptographic Accumulator:} }
Cryptographic Accumulator is an active research topic in cryptography. It describes a set of public keys with a short, verifiable signature. Given a public key, it can efficiently verify whether or not this key is a member of the group. In TBAC, we make use of a cryptographic accumulator to authenticate a \textbf{\textit{dynamic group}} of subject identities -- we do not need to define the \textit{exact} subject who is allowed to finally redeem the tokoin, but rather specify a \textit{group} of subjects. This is reasonable as we might not know who is the final subject when we mint a tokoin. 
As shown in our case study, the subject must be a licensed deliveryman but it is hard to know ahead of time the exact deliveryman who actually redeems the tokoin. Therefore we do not appoint an exact deliveryman when the tokoin is created; instead, we employ a crptographic accumulator to specify a group of licensed deliverymen who can be legal subjects to redeem a tokoin. Technically, a cryptographic accumulator is usually implemented by a strong one-way hash function that satisfies the commutative law. 

\subsection{Authentication, Authorization, and Auditing Processes}
\label{sec:AAA}

\subsubsection{Authentication}
\label{sec:Autentication}

The Authentication process verifies the authenticity of an identity or a tokoin capability in TBAC. In service of the Authorization process, the Authentication process in TBAC makes use of functions $\mathsf{Verify()}$ and $\mathsf{PolicyCheck()}$. 

{\bf Implementation of $\mathsf{Verify()}$.}
As mentioned in Section~\ref{sec:TBAC:definition}, $\mathsf{Verify()}$ is an overloading function that performs the tasks of verifying a message signature, a tokoin validity, or a participant's access privilege over a tokoin. To verify a digital signature, $\mathsf{Verify()}$ takes inputs $\sigma$ and $pk$, and outputs 1 if and only if $\sigma$ was indeed signed by $pk$ according to ECDSA. To verify the validity of a tokoin $t$, $\mathsf{Verify()}$ takes input $t$ and outputs $t.\textit{isValid}$. If $\mathsf{Verify()}$ takes inputs $t$, $pk$, and $\textit{role}$ i.e., to verify the role of participant $pk$ in $t$, we need to consider two different cases:
\begin{itemize}
\item if $\textit{role}=r_S$, i.e., $pk$'s subject role needs to be verified, $\mathsf{Verify()}$ needs to check the cryptographic accumulator $\mathsf{acc}$ stored in the access policy carried by $t$ and return 1 if and only if $pk$ is indeed a legitimate subject for $t$;
\item if $\textit{role}\neq r_S$, $\mathsf{Verify()}$ returns 1 if and only if $t.pk_O=pk$ if $\textit{role}=r_O$, or $t.pk_H=pk$ if $\textit{role}=r_C$.
\end{itemize}

$\mathsf{Verify}(t, pk, \textit{role})$ must be called by all the tokoin manipulation functions (except $\mathsf{Create}$) as \textit{role} defines the privilege of a participant over tokoin $t$. For example, $\mathsf{Redeem}$ should be called only by a legit subject while only the owner of a tokoin can modify the tokoin. Note that separating the verification of a tokoin holder and a legit subject is a good idea because we do not restrain the transfer of a tokoin, but we do have a clear definition on the subjects. In other words, there could be a case where the tokoin is transferred to a wrong participant who is not a subject, but we permit only legit subjects to access.


The implementation of $\mathsf{PolicyCheck()}$ requires a secure and reliable access control object $r_{\textit{ACO}}$ that samples the current environment to verify whether the access constraints are satisfied and whether the access procedure is strictly followed. It is essentially a trusted data intake source that is external to the blockchain system. 
In TBAC, we employ TEE to host $r_{\textit{ACO}}$ as it provides a trusted execution environment that physically isolates security-sensitive programs to prevent them from being tampered with. 

To fully implement $\mathsf{PolicyCheck()}$, we need the object $r_{\textit{ACO}}$ to do two things: to securely collect the $\textit{evidence}$ from the physical or digital environment upon receiving the tokoin redeem call, and to securely communicate with the blockchain system,  parse the tokoin to extract the access policy, and then compare the policy carried by the tokoin with the evidence collected from the environment to decide whether the access constraints are satisfied and whether the access procedure is strictly followed. 

Collecting evidence from the TEE environment is not trivial. A possible approach is to implement the drivers of the monitoring devices for evidence collection in the TEE secure zone and connect them directly to the external corresponding physical devices, to make sure that the devices can collect correct and authentic data without suffering from data interception or alteration. For example, in our case study, we implement the driver of a GPS receiver to verify the spatio-temporal access constraints and that of a camera to monitor the access procedure in a TEE hardware plugged into the sensor controller. Note that for a more complicated application, multiple controllers with the TEE hardware can be connected via secure protocols such as SSL/TLS to monitor a large environment. Under the security assumption that the TEE secure zone is physically isolated and tamper-free, one can safely accept the $\textit{evidence}$ as it is directly and securely captured. 

Next we implement the access control object $r_{\textit{ACO}}$ as a blockchain client inside the TEE secure zone to ensure that it can directly communicate with the blockchain network through a secure communication channel supported by existing techniques such as SSL/TLS. Such a design allows  $r_{\textit{ACO}}$ to work independently as a light-weight blockchain node that does not participate in ledger maintenance and consensus but can listen on the network layer messages without relying on any unreliable proxy. 
After securely fetching the access policy $t.\textit{policy}$ from tokoin $t$, $r_{\textit{ACO}}$ first checks whether the redeem request does come from a legit subject. This can be done by verifying whether the subject is a member defined by the $\textit{Accumulator}$ in $t.\textit{policy}$. Then $r_{\textit{ACO}}$ checks whether other access constraints such as the spatio-temporal constraints are satisfied by sampling the corresponding devices, and if yes, $r_{\textit{ACO}}$ grants the access by starting the access procedure, which first converts the actions defined in $t.\textit{policy}$ into instructions to make the resource available to the subject. During this procedure $r_{\textit{ACO}}$ keeps on communicating with the resource to be accessed and meanwhile sampling the monitoring devices to ensure that the access procedure is strictly followed. If any violation is detected during the access procedure, the access is  immediately stopped and appropriate measures are taken. It is all to these implementation considerations that make $r_{\textit{ACO}}$ securely verify whether the access constraints and procedure are met upon redemption. Note that all activities involved in $\mathsf{PolicyCheck()}$ are recorded in the TEE secure zone and will be logged in the blockchain as  the transaction script when $\mathsf{Redeem()}$ is completed.

\subsubsection{Authorization}

\begin{figure}[!htb]
	\centering
	\includegraphics[width=0.48\textwidth]{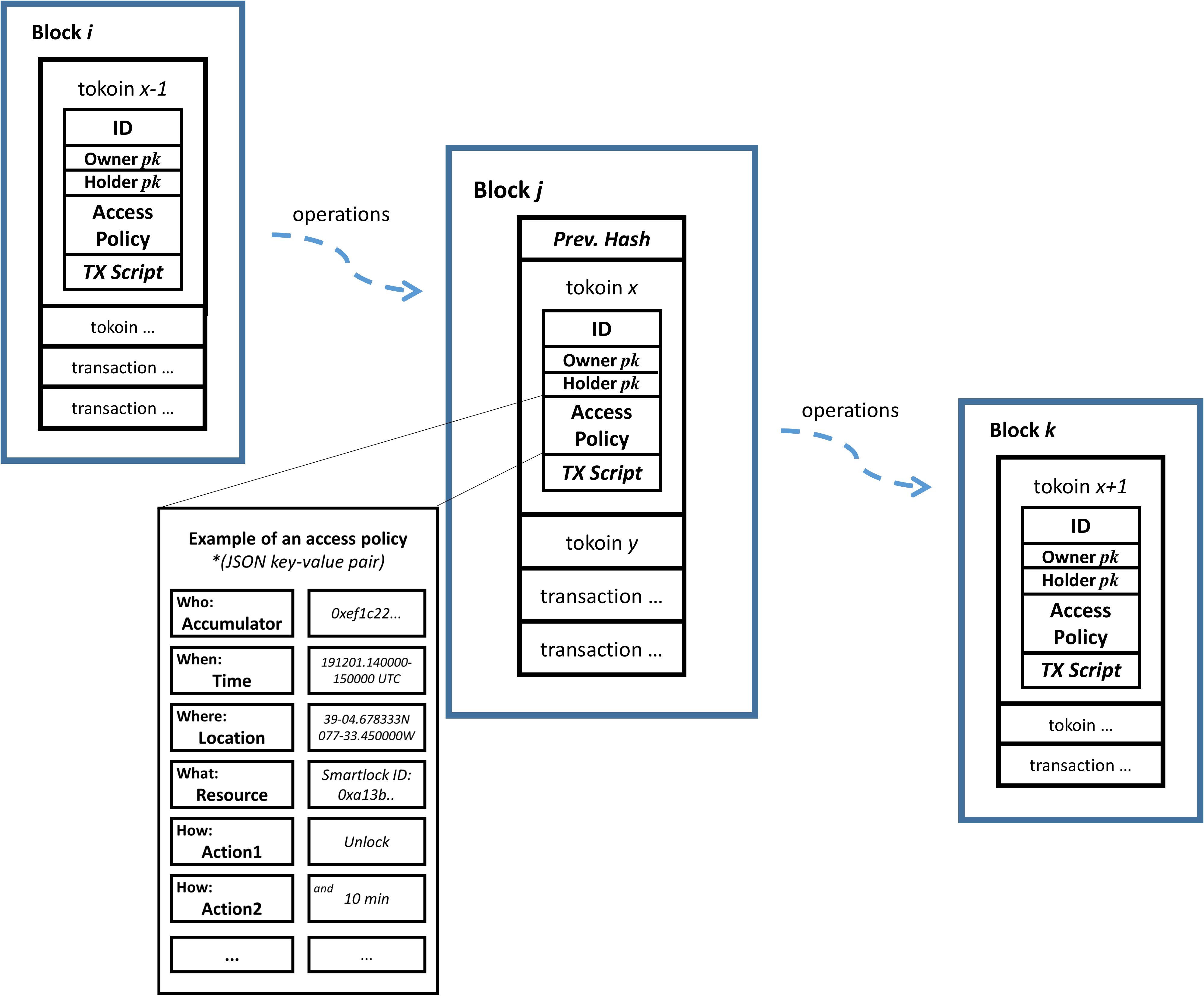}
	\caption{Block Structure and URPO}
	\label{fig:URPO}
\end{figure}

As mentioned earlier, an Authorization process determines whether a person or a process is authorized to perform a given access activity. In TBAC, this process grants a privilege to access the resource, modify the privilege, or redeem a tokoin under correct access policy. It includes the following five functions, whose implementations are detailed in sequel:
%
%
\begin{enumerate}
    \item $\mathsf{Create()}$: create a new tokoin;
    \item $\mathsf{Modify()}$: modify the access policy of an existing tokoin;
    \item $\mathsf{Transfer()}$: transfer a tokoin to another participant;
    \item $\mathsf{Revoke()}$: revoke a tokoin;
    \item $\mathsf{Redeem()}$: take in a tokoin and redeem the resource.
\end{enumerate}

The implementations of these five functions require explanation of the storage of tokoins. As said earlier, a tokoin should be stored securely and can be transferred atomically. 
%
To achieve this goal, we develop the UnRedeemed Policy Output (URPO) model, which is similar to Bitcoin's UTXO model, to manage the tokoins in the distributed ledger. Any operation over a tokoin, including the creation of the tokoin, starts from a message sent to all blockchain miner nodes for verification and consensus approval. If successful, the process takes in an existing tokoin if available, performs the operations as requested, and stores the processed tokoin in the next block within the ledger. This whole process is called a \emph{policy transaction}, and its output is either a newly created tokoin for $\mathsf{Create}$ or a modified one with the same $t_{\textit{ID}}$ and $pk_O$ as the input tokoin. As long as this policy transaction output is not redeemed or revoked, the tokoin remains valid ($t.\textit{isValid}=1$). 


The main purpose of this design is to maintain high atomicity and accountability of the tokoin operations, which implies that only atomic and one-to-one transitions can be allowed to operate a tokoin. Each transaction must be verified by all validating nodes in the blockchain with public knowledge, and no tokoin can be forged or forfeited out of thin air. The block structure is illustrated in Fig.~\ref{fig:URPO}. 
Each block contains a number of tokoins and each tokoin contains its native information as well as a $\textit{TX Script}$ field logging the corresponding message for and activities over the tokoin. An access policy within a tokoin is represented in the form of JSON key-value pairs for its simplicity \cite{XMListoolong,iampolicy}, as illustrated in Fig.~\ref{fig:URPO}. One can see that such a structure allows users to flexibly define their own access policies at different granularity levels. 


Any registered participant can create a tokoin, as long as it represents itself and issues access only to its own resource. To make a tokoin $\mathsf{Create()}$ call, the participant sends out a message $msg:[pk_O||t_{\textit{ID}}, \textit{create}, \textit{policy}]_{\sigma_{r_O}}$ signed by its secret key $sk_O$ to the blockchain system, where $t_{\textit{ID}}$ is designated by the participant and $pk_O||t_{\textit{ID}}$ is the composite ID of the created tokoin that is globally unique inside the blockchain. With the public key available, the miner nodes can $\mathsf{Verify}(\sigma_{pk_O}, pk_O)$, 
and create a new tokoin $t=(t_{\textit{ID}}, pk_O, pk_O, \textit{policy}, \textit{isValid=}1)$. Upon the creation of $t$, the participant acquires the role of $r_O$ with respect to $t$.  All actions performed during the creation of tokoin $t$ are logged in $\textit{TX Script}$ and recorded with $t$ in the next block as illustrated in Fig. \ref{fig:URPO}. 



The implementations of $\mathsf{Modify()}$ and $\mathsf{Transfer()}$ are similar to that of $\mathsf{Create()}$, in that they all require an authenticated request message carrying the corresponding op code sent to the blockchain system. But there are subtleties that differ them significantly: $\mathsf{Modify()}$ can be called only by a tokoin owner, and it changes the access policy of the tokoin; $\mathsf{Transfer()}$, on the other hand, can be called only by the current holder of a tokoin, and it carries the public key $pk_H^\prime$ of the next holder thus changing the current holder of the tokoin upon completion. Accordingly, the message for $\mathsf{Modify()}$ has a format of $[t, \textit{modify}, \textit{policy*}]_{\sigma_{r_O}}$ and that for $\mathsf{Transfer()}$ has a format of $[t, \textit{transfer}, pk_H^\prime, ]_{\sigma_{r_C}}$, for tokoin $t$.

When receiving a $\mathsf{Modify()}$ message, the verifiers in the blockchain system first check whether $\mathsf{Verify}(t, pk_O, r_O)=1$, and if yes, the access policy of the tokoin is changed and the corresponding transaction is recorded in the next block. Note that a revision on an access policy can modify any key-value pair of the policy, and can add new or delete existing key-value pairs, to redefine the access policy. Also note that the values within a policy are all plaintext modifiable except the access subject group, which consists of one or more individuals and is described by a cryptographic accumulator; therefore, to add or delete subject $pk^\prime$, ${\textit{Add/Del}}_{\textit{ACC}}(pk^\prime)$ and ${\textit{Update}}_{\textit{ACC}}()$ should be called to add or delete the subject and update the value of $\textit{Accumulator}$ in the access policy. Similarly, when receiving a $\mathsf{Transfer()}$ message, the verifiers first check whether $\mathsf{Verify}(t, pk_O, r_C)=1$, and if yes, the current holder of the tokoin is changed to $pk_H^\prime$ and the corresponding transaction is recorded in the next block. Note that $\mathsf{Modify()}$  and $\mathsf{Transfer()}$ do not create a new tokoin but revise certain field of an existing tokoin, and the revision actions are recorded in the $\textit{TX Script}$ field of the corresponding transactions. 


The implementation of $\mathsf{Revoke()}$ is rather simple. To revoke a tokoin $t$, the owner of $t$ sends out a message $msg: [t,  \textit{revoke}]_{\sigma_{r_O}}$ to the blockchain system, in which the verifiers first check whether $\mathsf{Verify}(t, pk_O, r_O)=1$, and if yes, $t.\textit{isValid}$ is set to 0 nullifying the tokoin $t$ in the system. Note that only the owner of $t$ can revoke $t$.


The implementation of $\mathsf{Redeem()}$ is a bit complicated. Upon receiving a $\mathsf{Redeem()}$ message $msg:[t, redeem]_{\sigma_{r_S}}$ from a subject $r_S$, the verifiers need to check $\mathsf{Verify}(t, pk_O, r_S)=1$, and if yes, $t$ is transferred to $r_{\textit{ACO}}$, who then calls $\mathsf{PolicyCheck()}$ to redeem the requested resource. If $\mathsf{PolicyCheck()}$ successfully returns, which means that the access process is successful, $r_{\textit{ACO}}$ invalidates the tokoin $t$  and then sends a confirmation message to the tokoin owner $r_O$. All activities in the redeem process, including those from $\mathsf{PolicyCheck()}$, are recorded in the $\textit{TX Script}$ field of the transaction for $\mathsf{Redeem()}$.

\subsubsection{Auditing}

Auditability and traceability are native in TBAC, as all operations over a tokoin and all resource access activities are logged within the $\textit{TX Script}$ field of a transaction stored in the blockchain. Such auditing evidence is publicized and verified by the whole blockchain system, and as a result, it is globally legit. Under the security assumption that the blockchain system is free from forking and the consensus process is not compromised, the auditability of tokoin manipulation and access control activities can be securely guaranteed.

\subsubsection{Summary}
There are three unique distinctions that differentiate TBAC from other access control models. First, fine-grained access control can be achieved by realizing an access policy that defines \emph{who is allowed to do what at when in where by how}, where \emph{who}, \emph{what}, \emph{when} and \emph{where} constitute the access constraints that must be satisfied in order for an access request to be granted and \emph{how} describes the access procedure that must be strictly followed to guarantee access safety while avoiding access violations. Second, an access privilege, which is intrinsically metaphysical, is transformed to a digital asset, i.e., a tokoin, that is stored in a blockchain and managed with atomic and accountable operations as if it is a cryptographic coin. Third, the validation of the access policy is performed at the TEE secure zone that physically protects all related programs and securely collects environment evidence for correct access decisions and access procedure monitoring. These design considerations ensure that TBAC can provide fine-grained and accountable access (procedure) control with cryptographic level security confidence while making unauthorized access impossible.  

Now one can safely claim that TBAC ensures all access activities to be securely restrained to what the resource owner has authorized to (\emph{no overprivilege}) and to be publicly recorded with no stealthy access (\emph{no weak auditability}). 

%

\subsection{TBAC Prototype Implementations}

Our TBAC system consists of three components: (1) a blockchain-based distributed ledger that securely manages the tokoin access capabilities, supports secure atomic tokoin operations in the form of transactions, and logs all activities with security significance for auditing purpose, (2) a trusted local access control system within a TEE chipset that hosts the programs of embedded blockchain clients and sensor drivers in its secure zone for collecting trusted environment evidence upon redemption and making correct, attack-free access control decisions, and (3) an App-based user interface for a good user experience.

We have two implementations of the first component: a native implementation \emph{Go-Tokoin} in Go language (Golang) that follows our original design for the best performance, and an Ethereum based implementation \emph{Ethereum-Tokoin} in Solidity that shows adaptivity of TBAC to mainstream blockchain platforms. We run Ethereum-Tokoin in both Ethereum Mainnet and Quorum, a consortium fork version of Ethereum that uses Raft or Istanbul-BFT consensus\footnote{https://github.com/jpmorganchase/quorum}. The tokoin functions are tested on the Native Go-Tokoin and the adaptive Ethereum-Tokoin. The experiments are run with seven virtual nodes on top of a PC with the following setup: 8-Core Intel i7-6700HQ @ 2.6GHz, 16G memory, Ubuntu 18.04.1 GNU/Linux. We record and analyze the performance of the implementations for the same case study in Section.\ref{sec:casestudy}. One can preview the results in Fig. \ref{fig:exec:time}, which indicate that our native Go-Tokoin takes typically 40-60 millisecond to confirm each transaction, while the Ethereum-Tokoin in Quorum consortium chain takes about 1 second (more than a magnitude) and that in Mainnet takes about 30 to 50 seconds (one more magnitude than that). 


\subsubsection{Go-Tokoin: Native Golang Implementation}


We implement the main blockchain system of TBAC with over 4000 line of codes in Golang, including the 
Tendermint-BFT consensus and the intercommunication protocols between a participant and the blockchain ledger. Golang is selected because it is a memory-safe, high-concurrent, high-usable language that is quite popular in the security community.  As we build our native system pretty much from scratch, we concentrate on flexibility and customized optimization while strictly following the detailed construction presented in Section~\ref{sec:TBAC:definition}. 
A complete working system is available in Github at https://github.com/zhuaiballl/Go-Tokoin.

\subsubsection{Ethereum-Tokoin: Adaptive to Mainstream Platforms}

Although our Go-Tokoin native implementation has better performance and more design flexibility as demonstrated by our case study in Section~\ref{sec:casestudy}, we still want TBAC to be readily available in other mainstream blockchain platforms. Thus we implement all required TBAC functions in Ethereum 
Solidity, in the form of a smart contract  (technically called interface). As we manage tokoins as digital assets, we develop the interface on top of Ethereum ERC token standards, which stand for Ethereum Improvement Proposals. Popular ERC token standards include ERC-20, ERC-777, etc.; however most of them create coins that are \textit{fungible}, which means that all the coins are identical and of the same data. This can perfectly serve the need of cryptocurrency, but clearly it cannot meet the need of TBAC, as we may modify and revoke a tokoin, and need to trace and audit all tokoins. Thus we choose ERC-721, a special ERC token standard that is often used to represent the different ownerships over digital assets or collectibles, and can be tracked individually. Then, we develop our own TBAC Smart Contract interface on top of ERC 721, implementing the aforementioned TBAC-specific functions, Ethereum events, and data members.
Each user can mint a tokoin by creating a tokoin smart contract. Users can directly implement their tokoin contracts with Remix-Ethereum IDE, or simply use the TBAC mobile App (presented in the following subsection) to auto-generate one. We keep this open-source on the Github at https://github.com/DES-PER-ADO/Ethereum-Tokoin. 

The implementation of Ethereum-Tokoin is harder than that of the native Go-Tokoin in two aspects. First, altering access policies in Ethereum-Tokoin is hard, as smart contracts do not allow data post-alteration whatsoever. We employ an engineering trick to decouple the logic and the data with two different contracts to mitigate this problem. Second, during the tokoin redemption process, we need to intake trusted data from our TEE counterpart and sensors. However, taking in exogenous, non-blockchain generated data from external sensors into a smart contract is cumbersome. This is because of the intrinsic exclusiveness and deterministic requirements of the smart contract environment. It is null to upload data from a single point to the smart contract, as the Ethereum VM requires the same logic and same data input at all nodes. To overcome this problem, we require $r_{\textit{ACO}}$ to post the sensor data through command \texttt{sendTransaction} to the full blockchain public network.

%
%

\subsubsection{The Robust Access Control Object}
The most popular hardware-assisted TEEs include Intel Software Guard Extensions (SGX), ARM Trustzone technology, and AMD Secure Execution Environment. In our implementation, we adopt the ARMv8-M TrustZone, which was introduced by ARMv6 around 2002 and then became popular due to its low-power consumption. It is comprised of three components, namely the secure zone, the non-secure callable interface, and the non-secure zone. The secure code in the secure zone has higher privilege and can access resources in both secure and non-secure zones, while the non-secure zone is isolated from secure resources and programs.

To implement the TEE-enabled access control object, we adopt LPC55S69-EVK, an ARM Cortex-M33 based microcontroller secured by the ARMv8-M TrustZone. This is because LPC55S69-EVK is one of the very few off-the-shelf chipsets that support the ARM Cortex-M33 architecture. Compared to the Cortex-A series, Cortex-M33 has a much lower cost (\$10 for Cortex-M33 compared to {\raise.17ex\hbox{$\scriptstyle\sim$}}\$200 for Cortex-A and {\raise.17ex\hbox{$\scriptstyle\sim$}}\$400 for Intel SGX), and its secure area and non-secure area can be precisely partitioned, providing strong flexibility. The cost we have to pay is the availability of a very little library for development, thus a majority of our codes are written by us in C and Assembly. Compared to the rich libraries of Inter SGX and Cortex-A with developer-friendly IDEs, it looks like we build the TEE system directly on a baremetal MCU. This part takes up about 6000 lines of code.

Recall that we require $r_{\textit{ACO}}$ to achieve two objectives: i) to securely collect information that can verify whether the access policy is strictly followed, and ii) to securely communicate with the blockchain system. The information that should be collected is application-specific, and is dependent on the access policy specified by the tokoin $t$. For example, if \emph{when} and \emph{where} are defined in $t.\textit{policy}$, one can employ a GPS receiver to realize the first goal by implementing its driver in the TEE secure zone and connecting the external GPS hardware with its driver directly without any intermediary. 
To realize the second goal, we implement a blockchain client inside the TEE secure zone to ensure that it can communicate directly to our blockchain network via a SSL/TLS secure communication channel. After fetching the access policy $t.\textit{policy}$, its contents are parsed, based on which correct information can be collected to verify whether the access constraints (\emph{who}, \emph{what}, \emph{when} and \emph{where}) are satisfied and the access procedure (\emph{how}) is strictly followed. If yes, the tokoin redemption process is successfully completed.  
The code for our robust access control object is application-specific as the access policy (condition and procedure) verification are dependent on particular applications, and the one for the in-home cargo delivery case study is available at https://github.com/DES-PER-ADO/TACO.

\subsubsection{TAP: the TBAC App}
To avoid users from being messed up with different programming languages, and more importantly, to provide a better user experience with no code exposure, we develop an easy-to-use TBAC mobile APP in Android, namely TAP, that integrates a TBAC interface with a script/contract wrapper. The purpose of TAP is to keep users from being exposed to Solidity, Javascript, Golang, or C code.

When a user logs in with its credentials (for example a username-password pair or a private key), TAP identifies its identity, searches for all active tokoins the user is associated with, and lists them in different categories based on the user's role as does by a wallet. If the user wishes to carry out an operation over a particular tokoin, the tokoin needs to be clicked and the required information needs to be input via the tokoin interface. The input information is then wrapped into Go-Tokoin scripts, or compiled by the \texttt{solc} Ethereum Javascript commands. 

We notice that TAP needs to take different steps for different blockchain platforms. For Ethereum-Tokoin, we need to initialize a contract, manage tokoins within an existing contract, or query the public on-chain data. To initialize a contract, a user can either implement its own contract or modify a default one and define the access policy at the administrative level. Next, the \code{solc} compiler compiles the created contract into an EVM-compatible Ethereum ABI, which is then wrapped into the Javascript command \code{.calcContract.new()} from the \code{Web3.js} library. To manage tokoins, the user can skip \code{solc} and directly wrap the parameters captured by the textbox and scroll-lists them into JS commands. These operations must be sent using the \code{sendTransaction()} command from the \code{Web3.js} library. To query the public on-chain data, \code{call()} is used. For Go-Tokoin, we do not need an overqualified virtual machine; instead, one can simply capture the parameters, wrap them in a \textit{msg} with different opcodes, and then send the message via \code{.sendTX}.
One can see that TAP improves the usability by 
providing a simple interface for the users to interact with TBAC. 
Similarly, the code of TAP is application-specific and the one for our in-home cargo delivery case study is available at https://github.com/DES-PER-ADO/TAP-Cargo-Delivery

\begin{figure}[!htbp]
	\centering
	\includegraphics[width=0.5\textwidth]{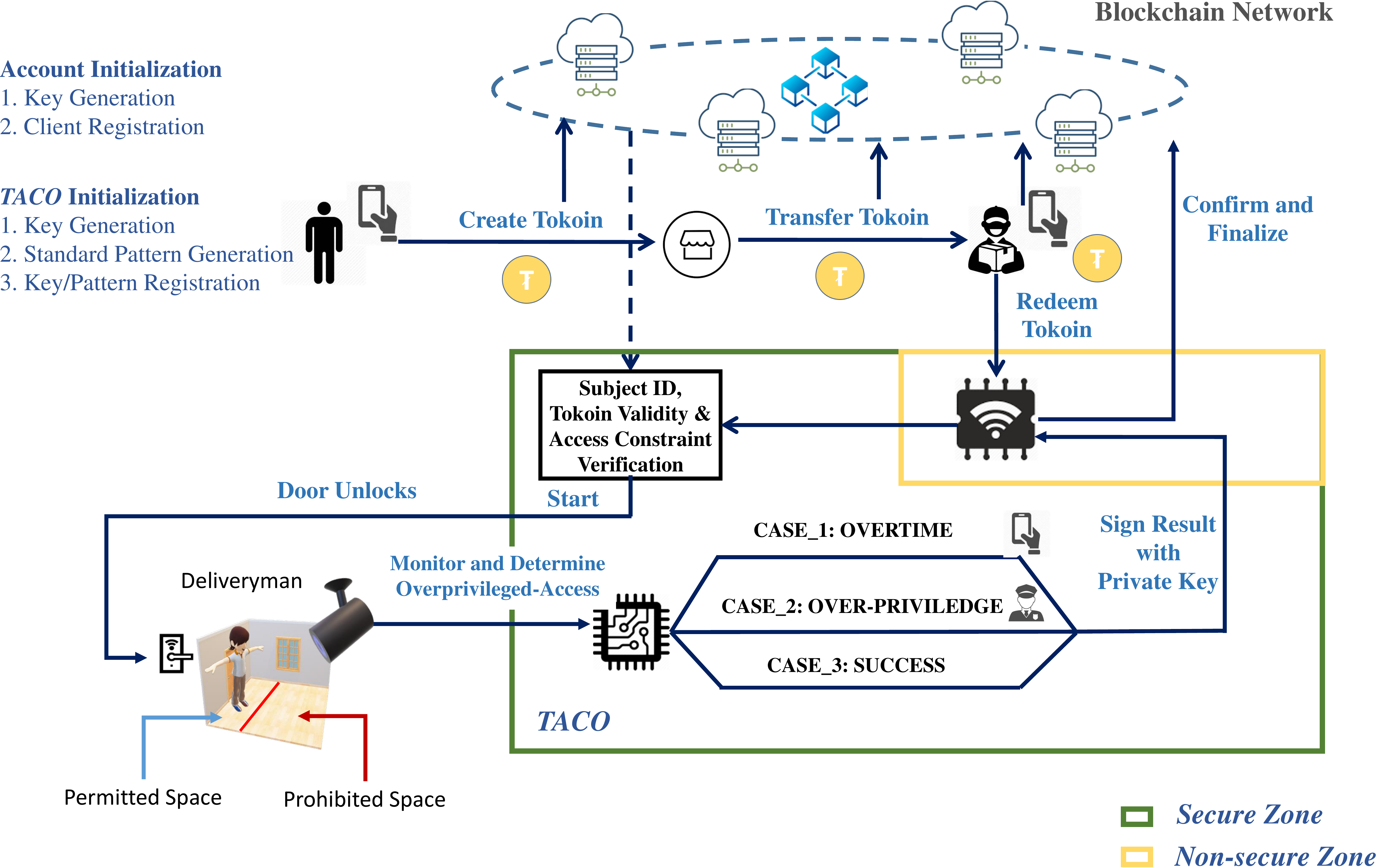}
	\caption{TBAC Assisted In-home Cargo Delivery}
	\label{fig:tokoin:cargo}
\end{figure}

\section{Case Study: TBAC-Assisted In-home Cargo Delivery}
\label{sec:casestudy}
In this section, we report a case study that employs TBAC to assist in-home cargo delivery. This case study demonstrates how TBAC and IoT can seamlessly work together to control a smart lock in a secure, fine-grained, and accountable manner for safe in-home delivery. 

In the US, online-purchased merchandises are usually delivered to the outside doorstep of a house, thereby risk of being stolen. With the help of the smart door lock, a deliveryman can open the house door and leave the cargo inside. This may seem to be a good solution and in fact it has been adopted by Amazon \cite{amazoninhome}. Nevertheless, by signing up for this in-home delivery service, users would surrender to the unlimited, unconditional, and unauditable accesses to their homes as an unlimited access token is issued from the door lock manufacturer's access server to Amazon after authorizing Amazon the access privilege to the door. 
It may get worse if the Amazon's server is compromised or the token is stolen or abused. 
In this section, we show that with TBAC, one can have high confidence that only the customer-approved accesses can take place, with a complete auditing. We also emphasize that with a fine-grained access policy specified by the customer, a robust access control object can monitor the delivery procedure to ensure that the deliveryman does not intrude the house by doing more than dropping the package. 

Fig.~\ref{fig:tokoin:cargo} demonstrates our TBAC-assisted in-home cargo delivery case study. When an order is placed, a tokoin specifying the detailed access policy to the customer's house is also created. The order and the tokoin are sent together to the seller,  who then transfers the tokoin to the first courier of the package when it leaves the warehouse. In transit the tokoin changes its holder when the package is handed to a different courier. The customer can monitor this process via TAP and can change the access policy based on the shipment status. When a courier arrives at the doorstep of destination house, the tokoin is redeemed and the package is dropped inside home if access policy verification succeeds. For security and safety, the access policy specifies that the deliveryman cannot walk out of the mud area to enter the main house. Thus a video camera is adopted to monitor the procedure and a violation is reported immediately when detected. In the following we present this case study in two processes: an initialization process and a delivery process.  

\subsection{Initialization}
The initialization process consists of customer account initialization and the robust access control object initialization. Note that we assume that the seller and its couriers are TBAC clients thus no extra work is needed here. For the ease of presentation, we name a TBAC access control object a \textit{TACO}.

\begin{figure}[!htbp]
	\centering
	\includegraphics[width=0.4\textwidth]{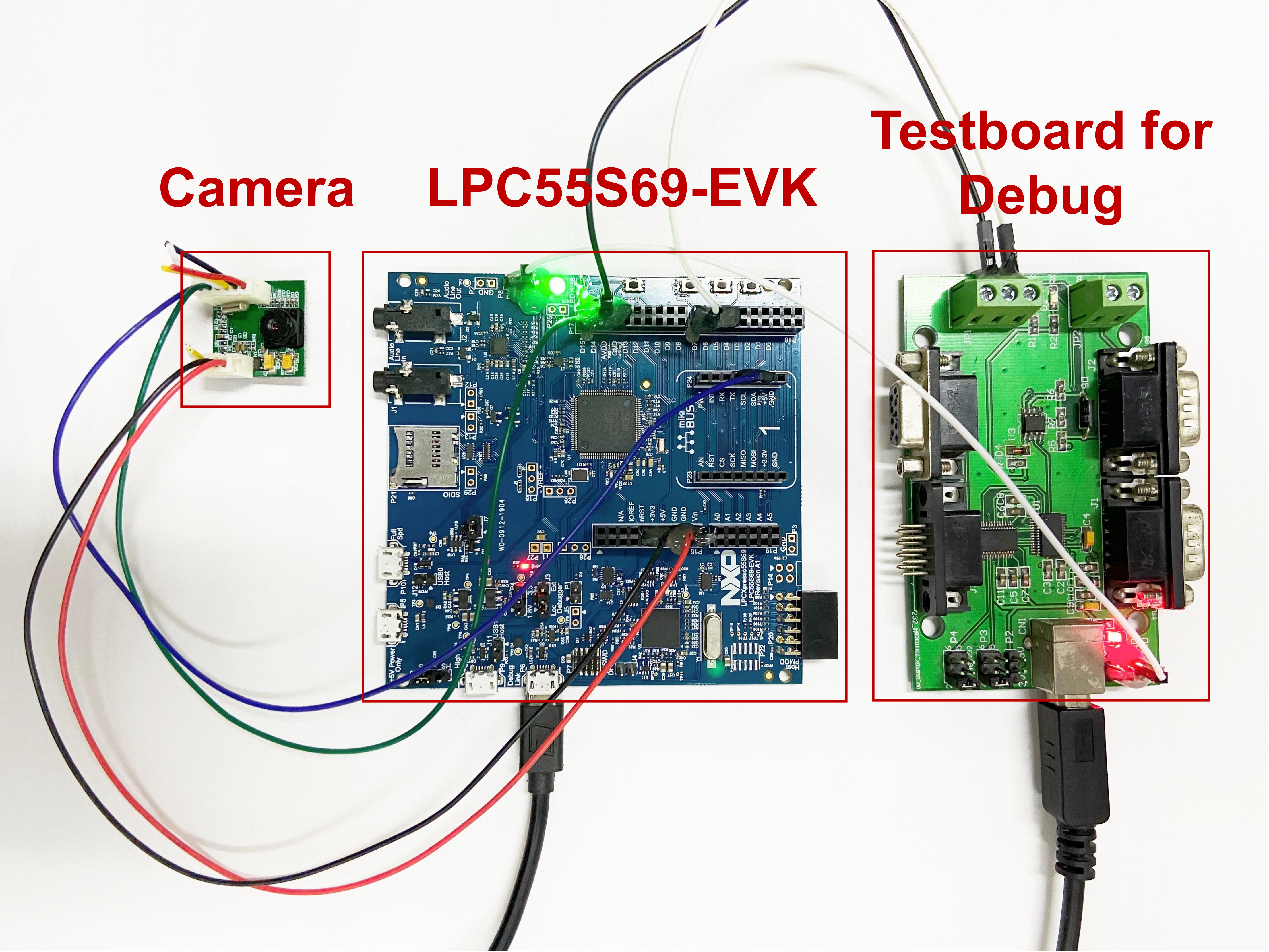}
	\caption{Experiment Setup of TEE}
	\label{fig:exp:TEE}
\end{figure}

\textbf{Account Initialization.} Upon initialization, a customer calls the \textsf{Gen} function to generate a pair of keys $sk$ and $pk$, registers itself with and broadcasts its public key $pk$ to the blockchain network, and keeps the private key $sk$ to itself.  The customer downloads the TAP software and becomes a TBAC participant who can talk with the blockchain securely. 

\textbf{\textit{TACO} System Configuration.} To establish the \textit{TACO} for our in-home cargo delivery case study, we adopt an LPC55S69-EVK microcontroller secured by an ARMv8-M TrustZone, a smart lock, a GPS receiver, and a low voltage UART serial image sensor camera. In addition, a testboard is utilized for debugging purpose. Upon initialization, \textit{TACO} generates a secret key $sk$ and a paired public key $pk$ using crypto-libraries in LPC55S69-EVK. Then it broadcasts its $pk$ to the blockchain for self-registration. The $pk$ is used as the address and the unique identifier of the \textit{TACO} system in the blockchain. \textit{TACO} also uses the camera to capture the \textbf{\textit{normal background}} of the home mud area when nobody shows up, and stores it as a \texttt{STANDARD PATTERN} into its secure zone for future detection of over-privileged behaviors such as the deliveryman walking out of the mud area to enter the main room. The $pk$ and the \texttt{STANDARD PATTERN} are then registered as a transaction in the blockchain. 

Note that we have implemented the drivers of a GPS receiver and a UART serial camera in the secure zone, which directly connects to the corresponding devices for secure data collection. Also in our case study \textit{TACO} is able to command the smart lock via secure communications with the lock server, who can securely talk with the smart lock, thus avoiding the hassle of writing a driver in the secure zone and connecting it to the lock. Fig.~\ref{fig:exp:TEE} illustrates the TEE setup with the sensor camera -- the GPS receiver is omitted for better illustration.

\subsection{Delivery Process}

The delivery process is depicted in Fig.~\ref{fig:tokoin:cargo}, whose steps are detailed as follows.

\begin{figure*}[!htb]%
	\centering
	\subfloat[Creating tokoin] {{\includegraphics[width=0.28\textwidth]{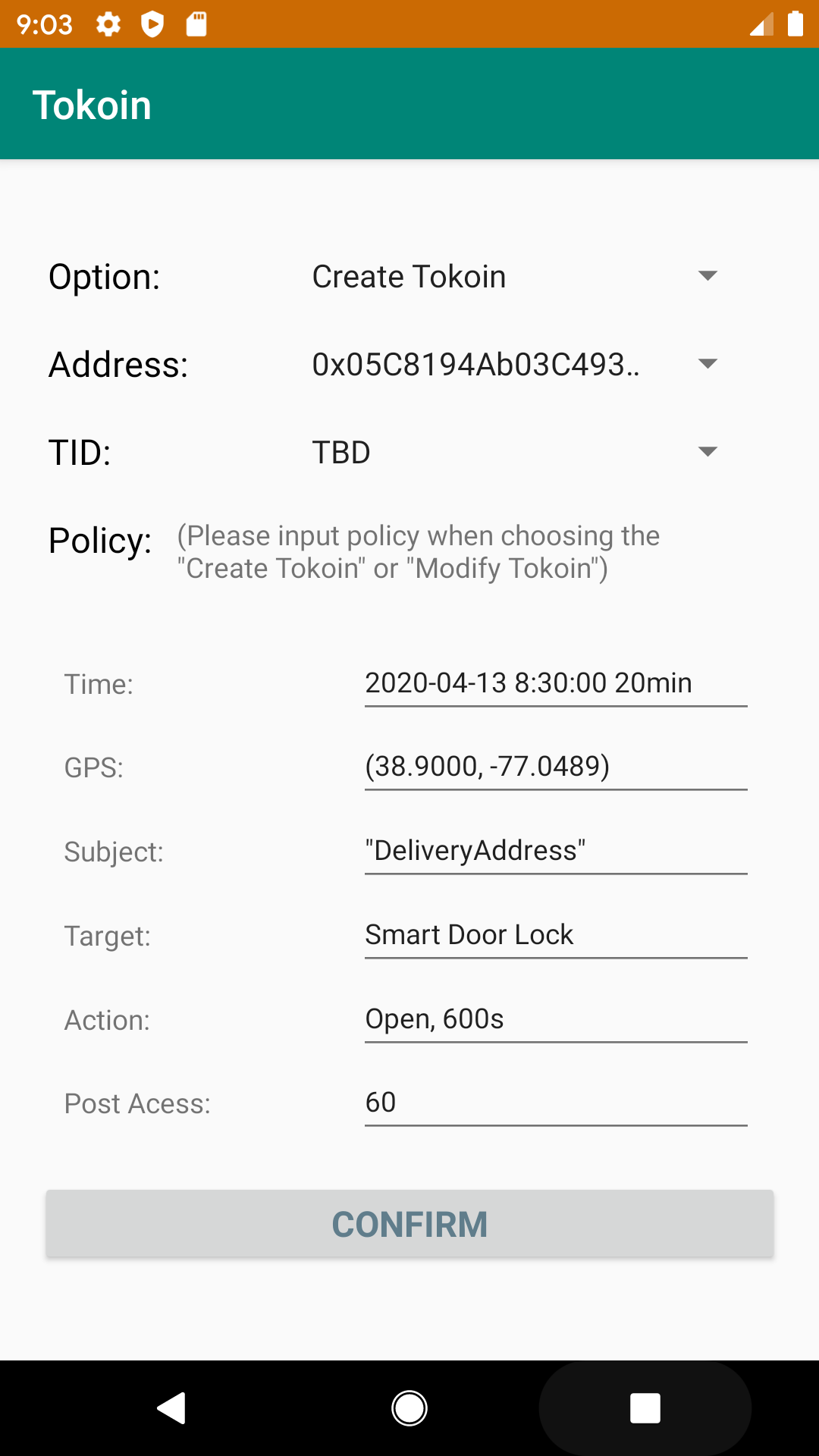} }}%
	\qquad
	\subfloat[Detailed information of a tokoin]{{\includegraphics[width=0.28\textwidth]{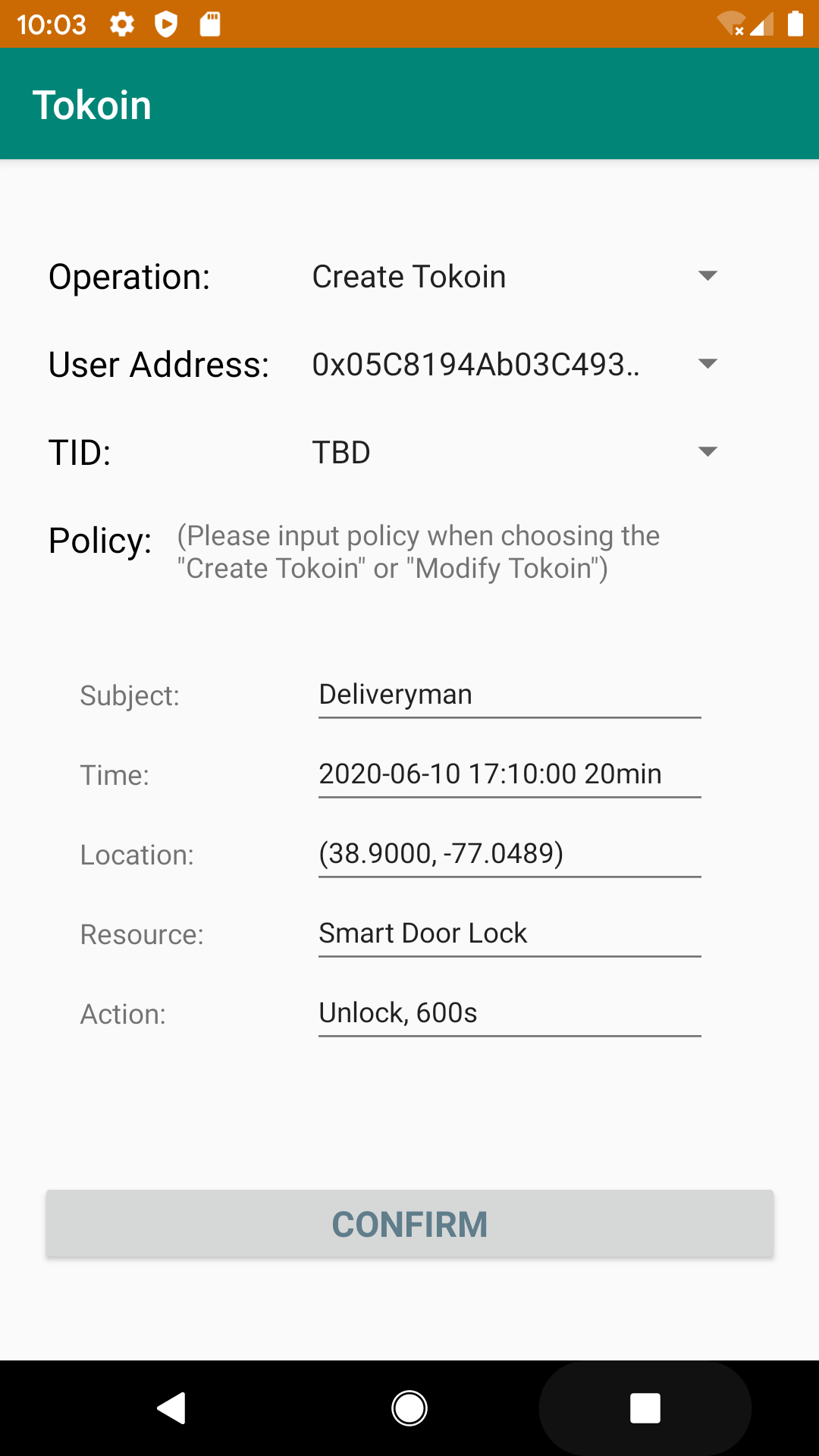} }}%
	\qquad
	\subfloat[Geographic trace of a tokoin]{{\includegraphics[width=0.28\textwidth]{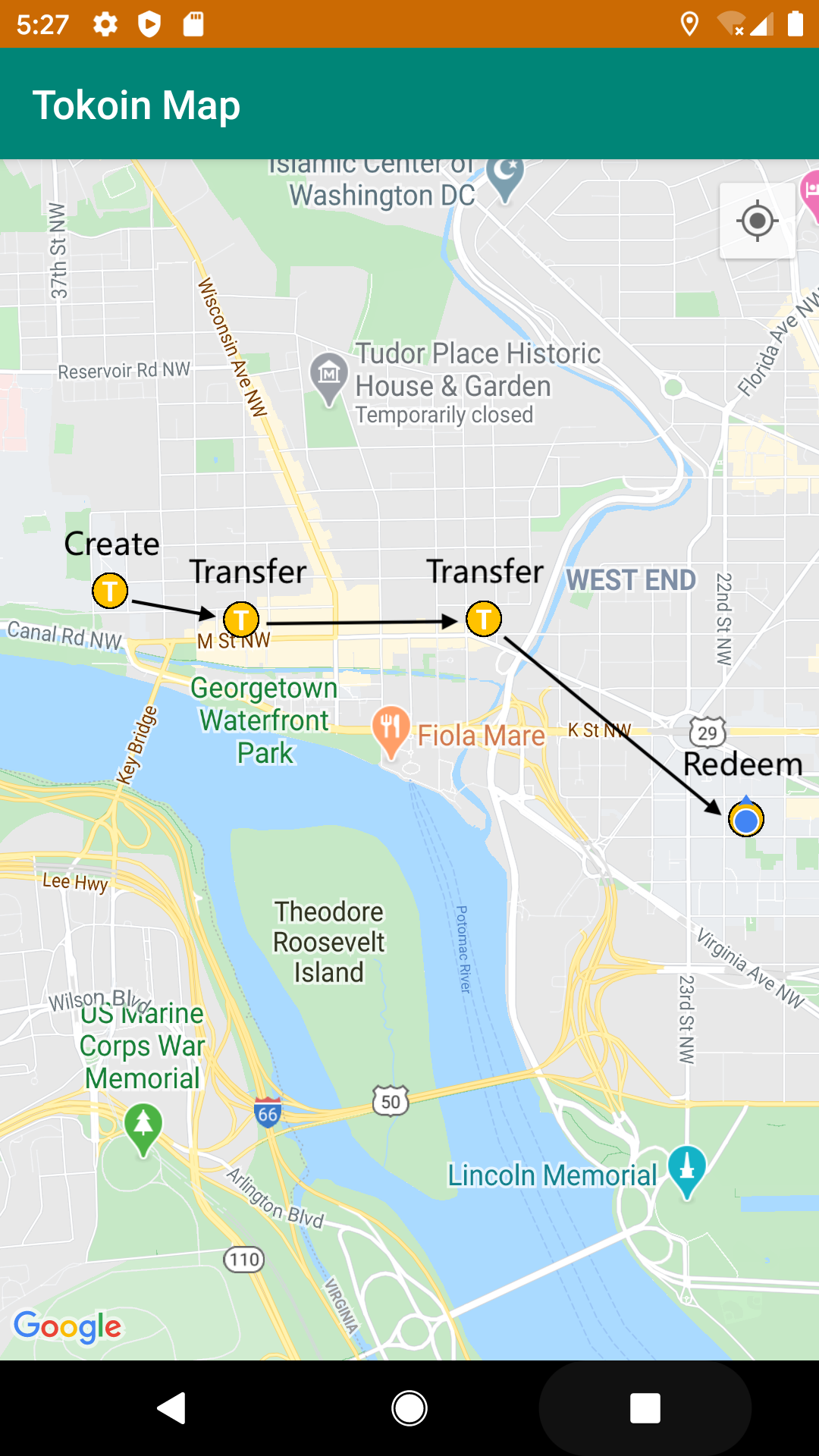} }}%
	\qquad
	\caption{Android Tokoin Activities for Cargo Delivery}%
	\label{fig:tokoinAndroid}%
\end{figure*}

\textbf{Create a tokoin.} Before an order is placed, the customer mints a tokoin by sending $msg:[pk||t_{\textit{ID}}, \textit{create}, \textit{policy}]_{\sigma_{r_O}}$ to the blockchain, who then creates a tokoin $t$ and logs it in a transaction. The $\textit{policy}$ specifies \textit{who} (any legit deliveryman) is allowed to do \textit{what} (cargo delivery) in \textit{where} (address of the house) at \textit{when} (e.g., 2-3PM tomorrow) by \textit{how} (enter the house, drop the package in the mud area, then leave the house in 5 minutes). See Fig.~\ref{fig:tokoinAndroid}(a) and Fig.~\ref{fig:tokoinAndroid}(b) for an illustration. The tokoin $t$ and the order together are then sent to the seller.

\textbf{Transfer the tokoin.} The seller then assigns a courier to this order, and transfers $t$ to the courier via a $\mathsf{Transfer}$ message when the package is handed to the courier. As we discussed before, $t$ can be freely transferred among the couriers through a standard transfer operation, which eases the re-distribution of the delivery job for better logistic convenience. Besides, the customer has the right to track the tokoin through a tokoin map as shown in Fig.~\ref{fig:tokoinAndroid}(c). It can also modify the tokoin (e.g., changing the delivery time window) during this process at its will before the tokoin is redeemed. 

\begin{figure*}[!ht]%
	\centering
	\subfloat[ An actual frame of a \textit{TACO} camera capture] {{\includegraphics[width=0.19\textwidth]{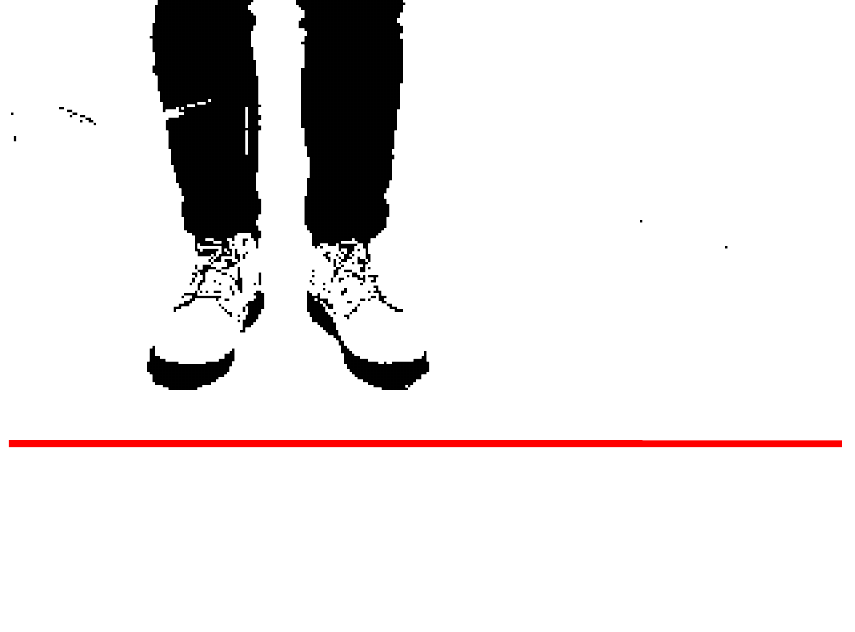} }}%
	\qquad
	\subfloat[ A  \textit{TACO} captured pattern of benign behavior ]{{\includegraphics[width=0.196\textwidth]{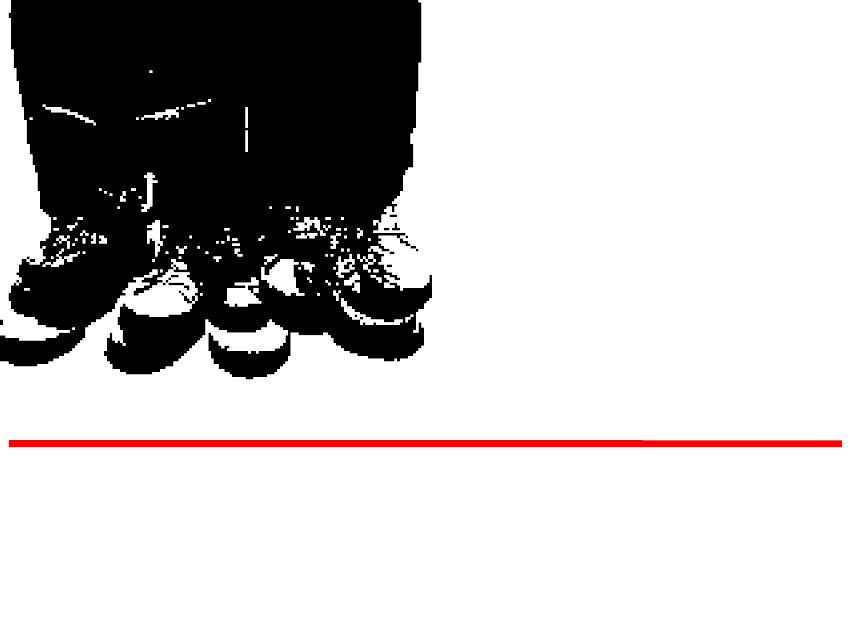} }}%
	\qquad
	\subfloat[ A \textit{TACO} captured pattern of minor violation behavior ]{{\includegraphics[width=0.196\textwidth]{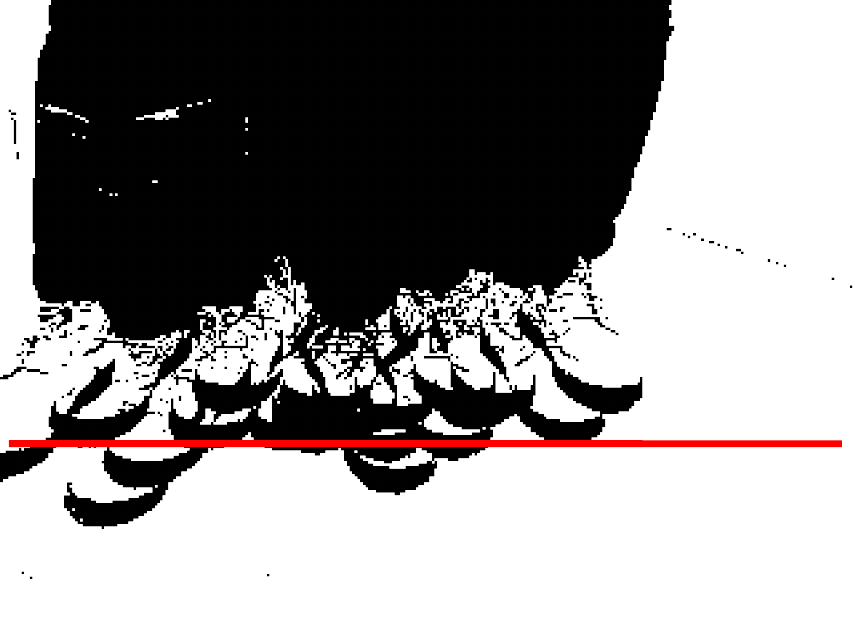} }}%
	\qquad
	\subfloat[ A \textit{TACO} captured pattern of a major violation behavior ]{{\includegraphics[width=0.196\textwidth]{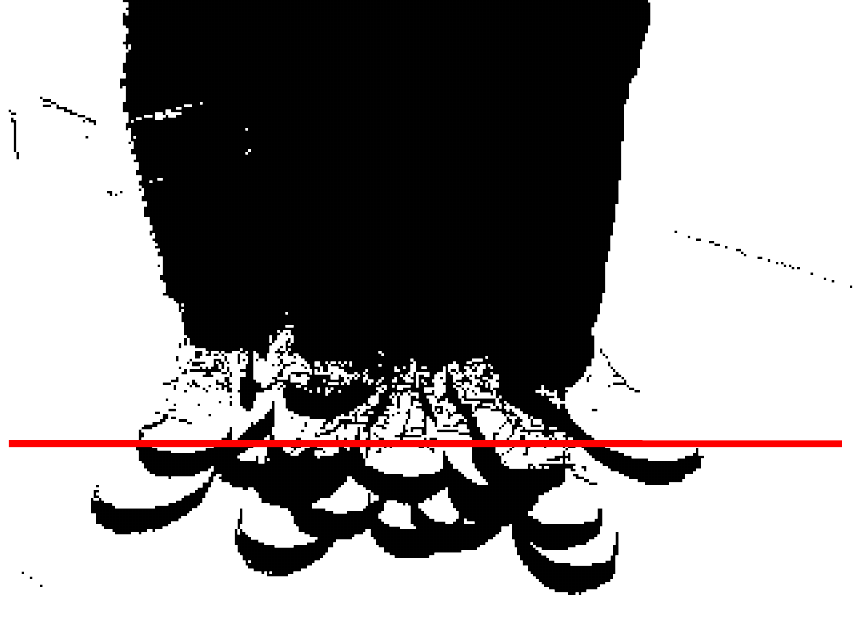} }}%
	\qquad
	\caption{A Set of Actual Captures of \textit{TACO} for Overprivileged-Access Monitoring}%
	\label{fig:TEEcapture}%
\end{figure*}

\textbf{Redeem the tokoin.} A deliveryman arrives at the house address and wishes to redeem the tokoin $t$ to complete the job. $\mathsf{Redeem}$ is then called, which sends the tokoin $t$ to \textit{TACO}, who would perform the following tasks for tokoin redemption.

\textit{Access Condition Verification.} After receiving $t$, \textit{TACO} needs to $\mathsf{Verify}$: 1. whether $t$ is a valid tokoin; 2. whether the deliveryman is a legit subject according to the cryptographic accumulator carried by $t.\textit{policy}$; and 3. whether the spatio-temporal access conditions are met, i.e., the time of delivery and the delivery address are all consistent with what are specified by $t.\textit{policy}$. If all verifications succeed, \textit{TACO} instructs the smart door lock to open the door and let the deliveryman in to drop the package. 

\textit{Access Procedure Monitoring.} After entering the house, the deliveryman  should drop the package in the mud area and leave in time specified by $t.\textit{policy}$. To monitor this process, \textit{TACO} constantly reads inputs from the UART serial camera and checks the position of the deliveryman by detecting moving objects in the video. Specifically, \textit{TACO} computes the difference between the \texttt{STANDARD PATTERN} and every video frame, and adds them up as a differential monitoring pattern, 
which is obviously a bitmap with boolean 1 for presence and 0 for absence of the deliveryman. The determination of a violation, i.e., an overpriviledged access, is detected if there is a boolean 1 out of the mud area, see the illustration in Fig.~\ref{fig:TEEcapture}, which uses an imaginary red line to represent the boundary of the mud area. 

There are two possible types of violations:
\begin{itemize}
	\item{Case 1:} the deliveryman stays longer  than the time specified by $t.\textit{policy}$. In this case, \textit{TACO} would first ring an alarm bell, then send a signed \texttt{OVERTIME} message to the customer. If the deliveryman does not leave immediately, \textit{TACO} may call the police.
	
	\item{Case 2:} the deliveryman walks out of the permitted mud area to enter the main room.  If this case is detected, as shown Fig.~\ref{fig:TEEcapture}(c)(d), the corresponding pattern of motion trajectory is recorded as a proof of an over-privileged behavior; then \textit{TACO} sends a signed \texttt{OVER-PRIVILEGED PATTERN} to the blockchain and takes appropriate measures such as locking the smart door and calling the police.

\end{itemize}

\emph{Post-Access Management.} If no violation is detected, \textit{TACO} sends a signed \texttt{SUCCESS} to the blockchain after the deliveryman successfully drops the package in the mud area and leaves the house. Note that all the data for access condition verification and access procedure monitoring must be signed with the private key of \textit{TACO} and stored within the TEE secure zone. This can guarantee the integrity and the trustworthiness of the data. The data itself or a digest of the data (if the data is too big) is also sent to the blockchain to be included in the transaction as a script for auditing the \textsf{Redeem} operation. 

\begin{figure}[!htb]
	\centering
	\includegraphics[width=0.5\textwidth]{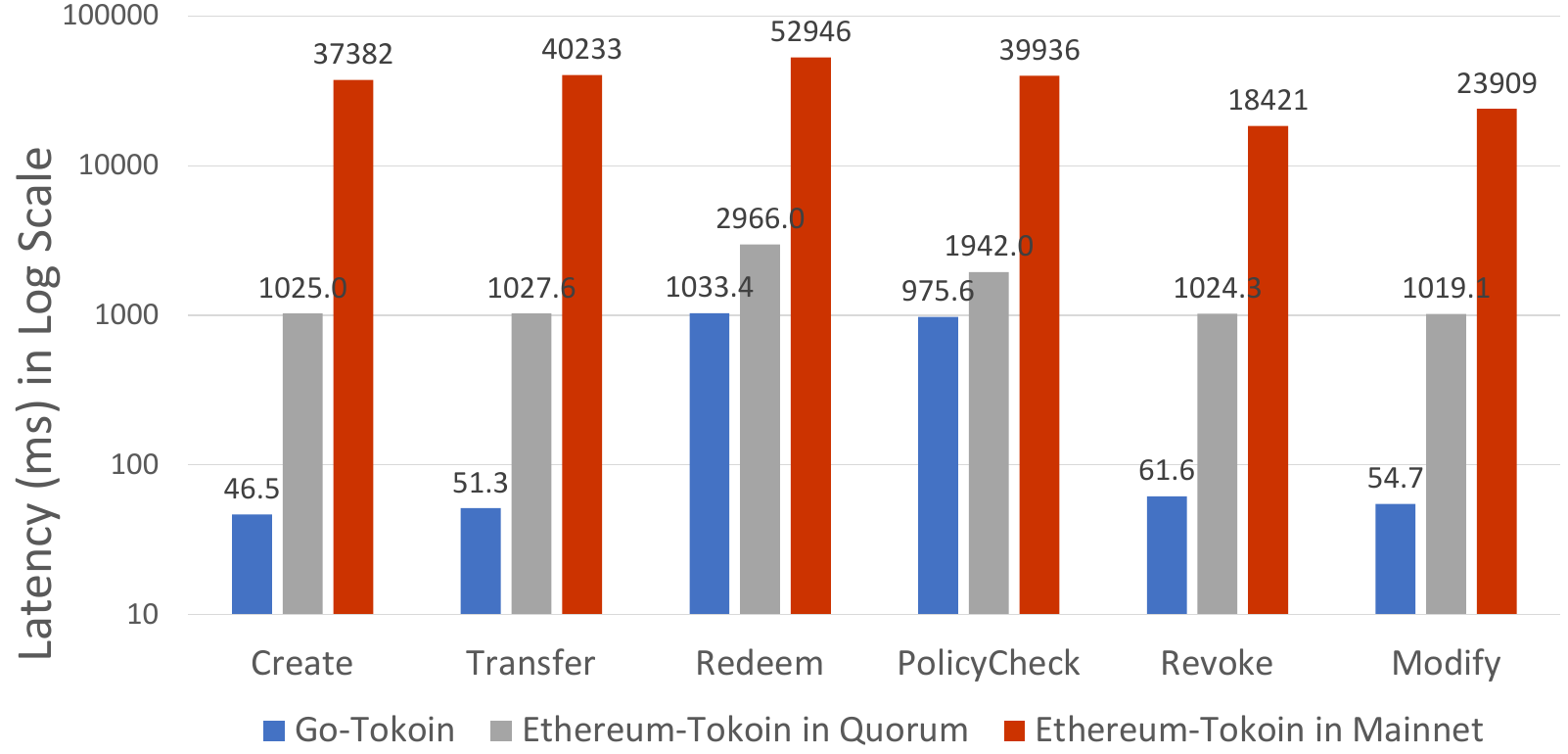}
	\caption{function Execution Times}
	\label{fig:exec:time}
\end{figure}

The total time for each tokoin manipulation function is reported in Fig.~\ref{fig:exec:time}. This figure shows the performance of different TBAC  implementations, in logarithmic scale. The reason why we choose log scale is because the confirmation times of Go-Tokoin and Ethereum-Tokoin in three different platforms vary up to two magnitudes. Our native Go-Tokoin takes typically 40-60 milliseconds to confirm each transaction, while the Ethereum-Tokoin in Quorum consortium chain takes about 1 second (more than a magnitude) and that in Mainnet takes about 30 to 50 seconds (one more magnitude than that). One should notice that Go-Tokoin takes much longer time in $\mathsf{Redeem}$, which includes the time for $\mathsf{PolicyCheck}$. This is because $\mathsf{Redeem}$ requires \textit{TACO} to sample the sensor readings, analyze the data, and take corresponding actions if needed, and communicate the data (or its digest) back to blockchain. Note that the time for monitoring the access procedure is excluded. 
Besides, the transaction cost on the Ethereum Mainnet is about 2967k Gas on average (\$14.7 USD, in July 16, 2019), while our consortium Go-Tokoin is free.

\subsection{Discussions on TEE Cost}
We notice that there exist other works that use TEE to perform trustworthy operations. However, most of them are prohibitively expensive, mainly because they use the more-easy-to-implement but expensive chipsets such as the Intel SGX or the Cortex A-series high-performance chipset. Such chipsets are mostly priced around \$300 {\raise.17ex\hbox{$\scriptstyle\sim$}} \$400, which is not practical for general systems used in our daily life. Cortex M-series chipsets are specifically designed for embedded systems and real time responses. This series has very few usable libraries, kernels, and operating systems, which poses a great engineering challenge on us. We build the system almost from the bare metallevel. To the best of our knowledge, we are the first to implement such a trustworthy system using the challenging yet cheap Cortex M23 MCU, which is priced around \$10.

\section{Conclusions and Future Research}
\label{sec:conclusion}

In this paper we propose TBAC, an accountable access control model that makes use of blockchain and TEE technologies to realize its goals of offering fine-graininess, strong auditability, and access procedure control. The basic idea of TBAC is to mint a tokoin that materializes the ``virtual'' access right into a cryptographically secure digital asset such that the resource owner can take full control of the access to its resource, without the need of delegating or relying on any third party such as a server. The access constraints and the access procedure as well as their contextual relationships can be precisely defined in a tokoin, which constitutes the access policy, thus realizing fine-graininess and access procedure control.  Moreover, all actions over the tokoin, either on-chain or off-chain, can be securely logged, facilitated by the blockchain and a robust access control object (implemented within the TEE secure zone) for policy compliance verification,  realizing strong auditability. We also present a TBAC-assisted in-home cargo delivery case study to demonstrate the effectiveness of TBAC in securing the procedure for a deliveryman to open a door and drop the package in the mud area of the room without entering the main room. Our TBAC design and the case study demonstrate a remarkable idea of extending trust from on-chain to off-chain, which has its own significance with broad applications and will be further investigated in our future research.

On the other hand, a tokoin in TBAC can be securely transferred and audited through a standardized process. This enables a new paradigm of trustworthy resource management such as secure IoT resource rental, trading of IoT resource right-of-use, or flexible errand delegation. For example, a senior executive may need a secretary to rebook a flight, change a hotel reservation, or reply a specific email on its behalf. The current solution is to provide the username/password pair or a picture of the credit card to the secretary. This certainly harms privacy and is undoubtedly highly insecure. If TBAC is properly applied, the executive can issue a tokoin to precisely specify the permitted operations without disclosing username/password or credit card information. This motivates us to investigate application-oriented TBAC implementations to address various domain-specific challenges in our future research.

\section*{Acknowledgments}
It was partially supported by the National Key R\&D Program of China under grant 2019YFB2102600, the National Natural Science Foundation of China under grants U1811463, 61971014, 61832012, 61771289, and 11675199, and the National Science Foundation of the US under Grants IIS-1741279 and CNS-1704397.

%
%

\ifCLASSOPTIONcaptionsoff
  \newpage
\fi



%

\bibliographystyle{unsrt}
\bibliography{liu.bib}

%



\begin{IEEEbiography}[{\includegraphics[width=1in,height=1.25in,clip,keepaspectratio]{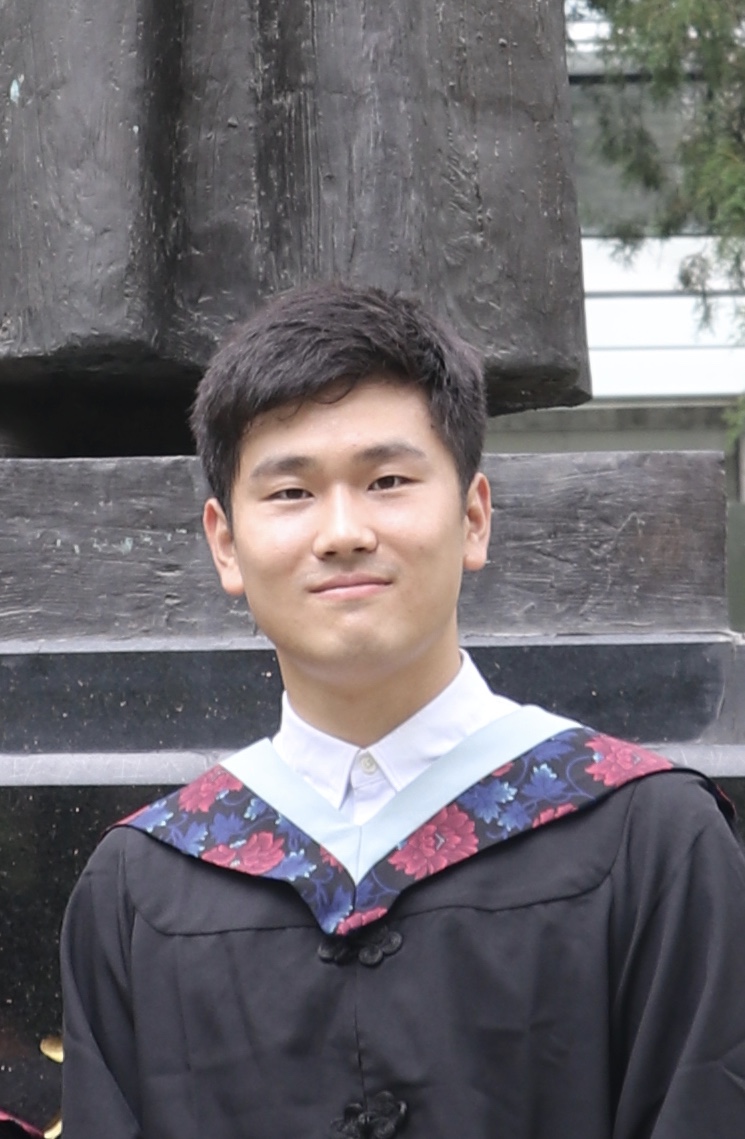}}]{Chunchi Liu} obtained his PhD degree from The George Washington University, Washington DC, USA, in 2020, and his BS degree with distinction from Beijing Normal University, Beijing, China, in 2017, both in Computer Science. His current research focuses on Blockchain, Internet of Things, Security, and Applied Cryptography. 
\end{IEEEbiography}

\begin{IEEEbiography}[{\includegraphics[width=1in,height=1.25in,clip,keepaspectratio]{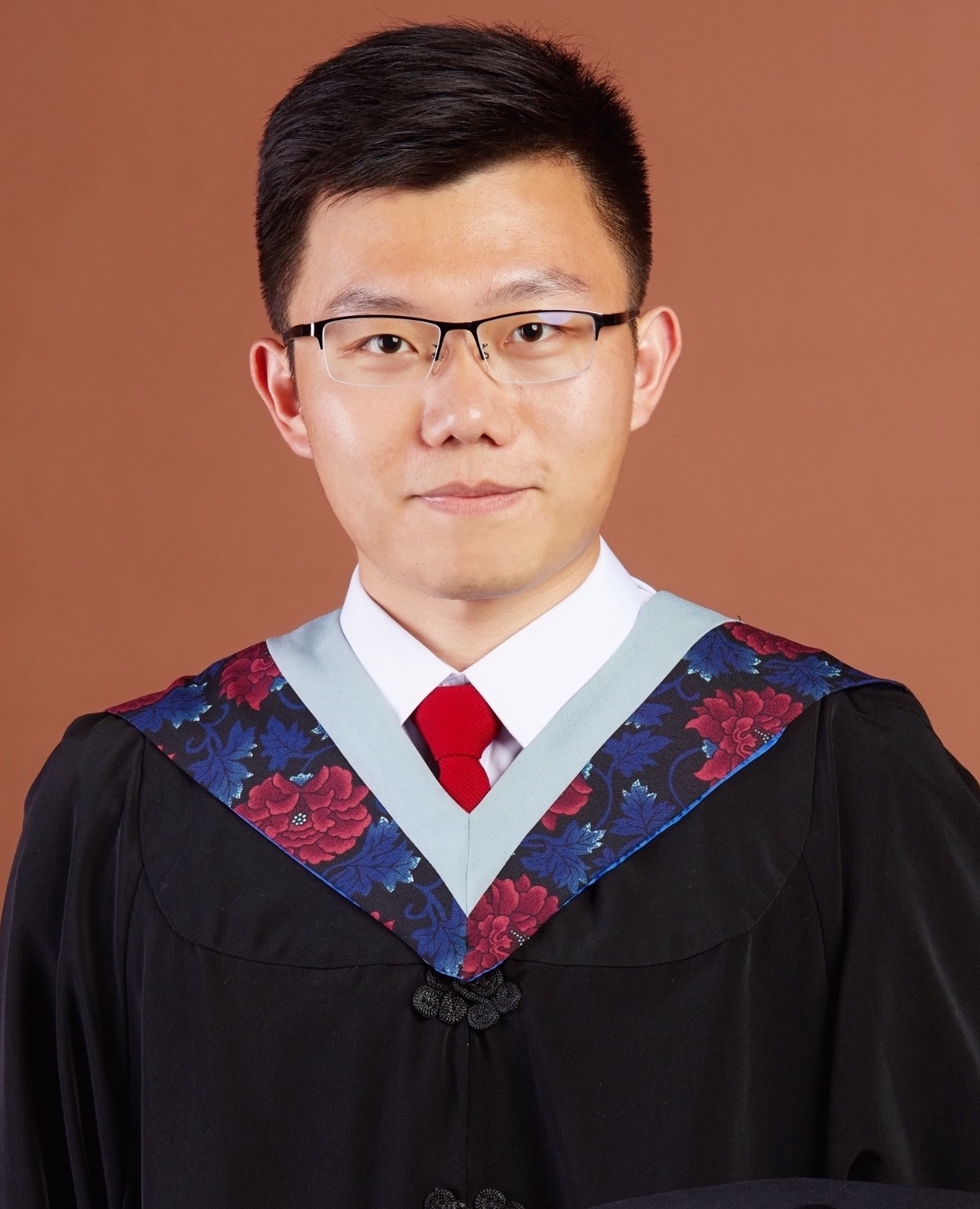}}]{Minghui Xu} is a PhD student in Computer Science at The George Washington University, Washington DC, USA. He received his BS degree in Physics in 2018 and minored in Computer Science during 2016-2018 from Beijing Normal University, Beijing, China. His current research focuses on distributed computing, blockchain, and quantum computing. 
\end{IEEEbiography}

\begin{IEEEbiography}[{\includegraphics[width=1in,height=1.25in,clip,keepaspectratio]{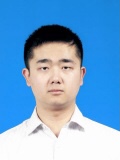}}]{Hechuan Guo} is a PhD student in Computer Science at Shandong University, Qingdao, China. He received his BS Degree in Computer Science in 2017 and MS degree in Engineering in 2020, both from Beijing Normal University, Beijing, China. His current research focuses on Blockchain, Consensus Protocols, Security, and Applied Cryptography. 
\end{IEEEbiography}

\begin{IEEEbiography}[{\includegraphics[width=1in,height=1.25in,clip,keepaspectratio]{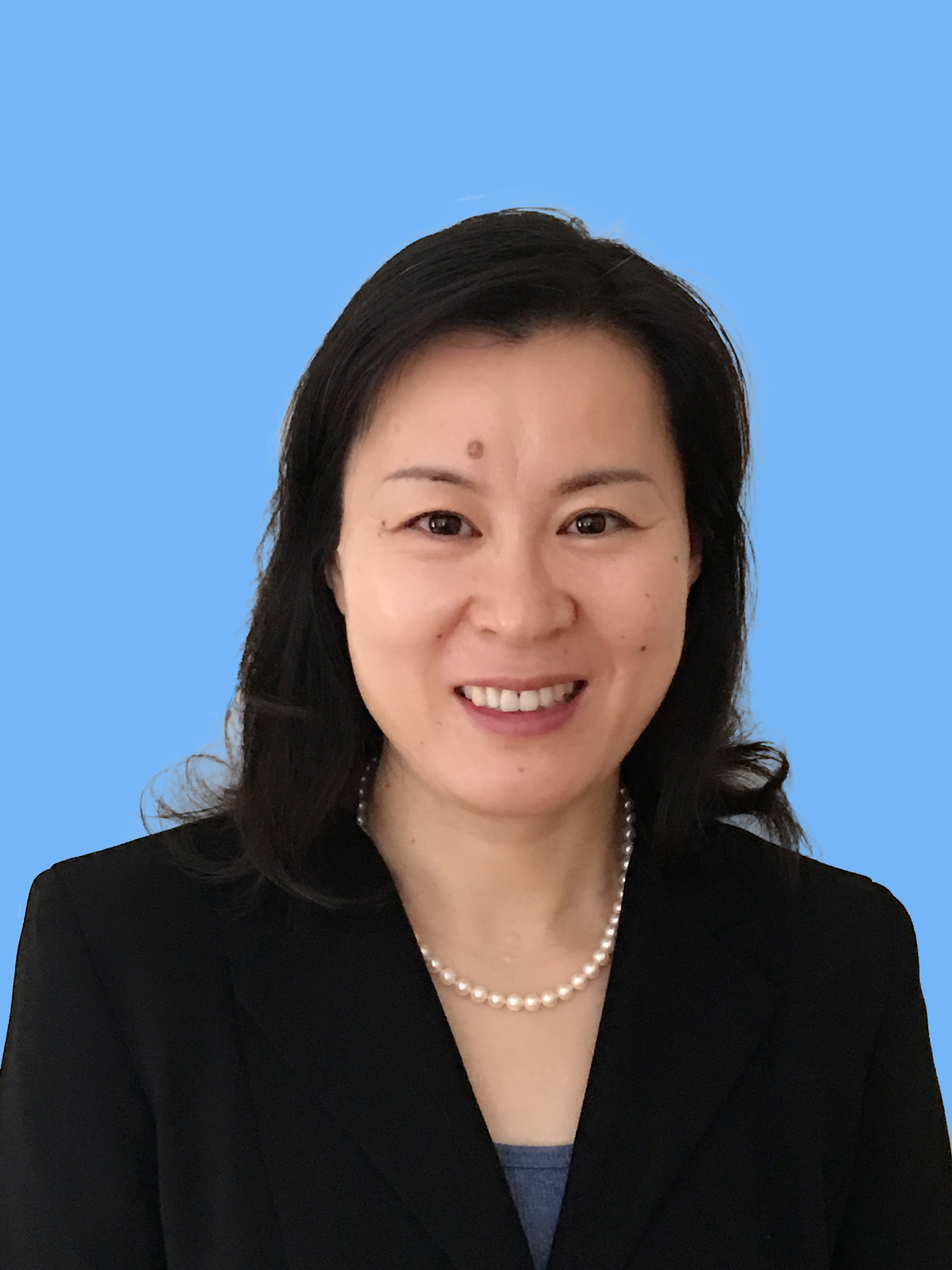}}]{Xiuzhen Cheng} received her MS and PhD degrees in computer science from University of Minnesota – Twin Cities, in 2000 and 2002, respectively. She was a faculty member at the Department of Computer Science, The George Washington University,  from 2002-2020. Currently she is a professor of computer science at Shandong University, Qingdao, China. Her research focuses on blockchain computing, IOT Security, and privacy-aware computing. She is a Fellow of IEEE.
\end{IEEEbiography}

\begin{IEEEbiography}[{\includegraphics[width=1in,height=1.25in,clip,keepaspectratio]{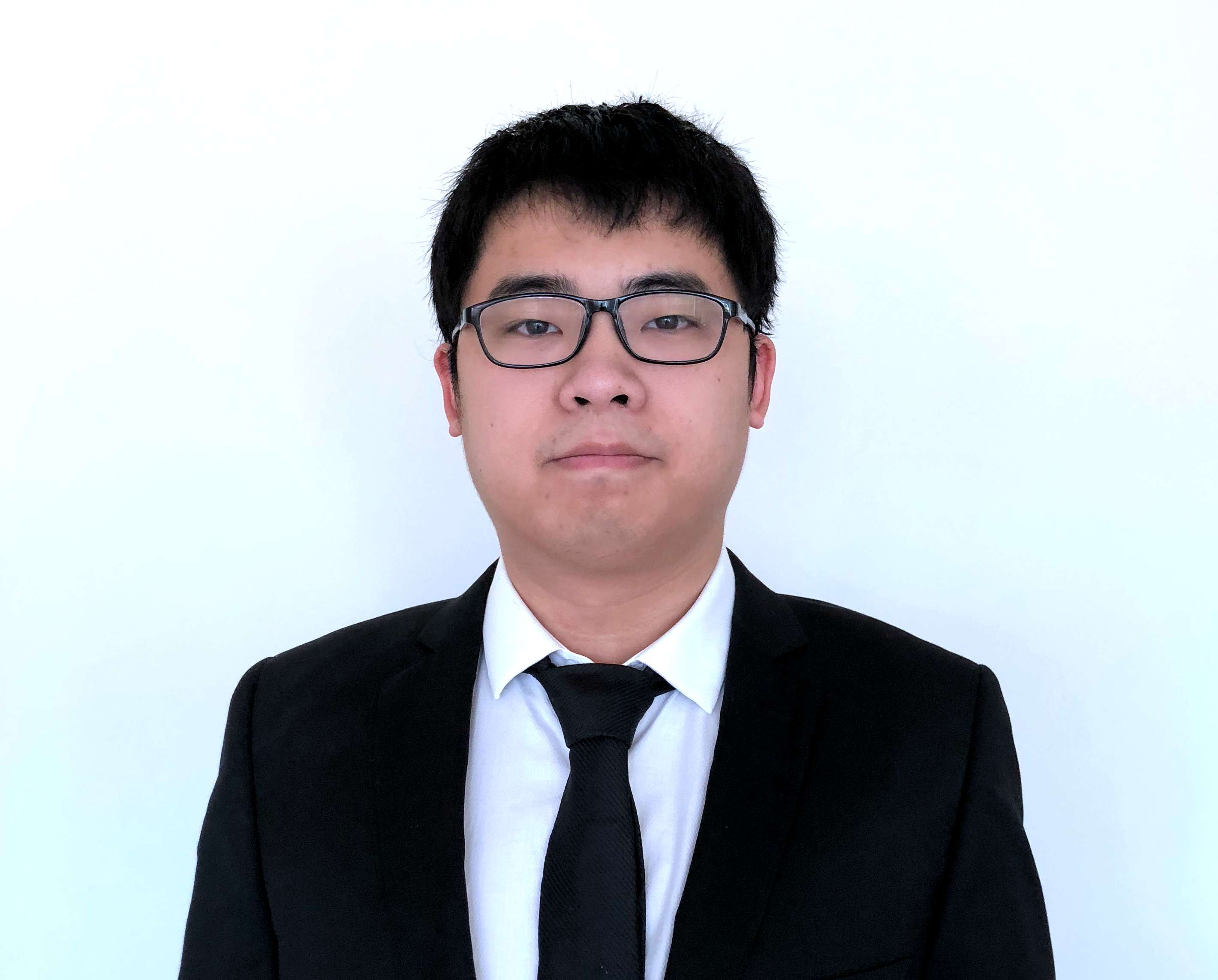}}]{Yinhao Xiao} received his Ph.D. degree in  computer science from The George Washington University, Washington, DC, USA, in 2019. He is a faculty member with the School of Information Science, Guangdong University of Finance and Economics, Guangzhou, China. His current research interests include IoT security, smartphone security, and binary security. 
\end{IEEEbiography}

\begin{IEEEbiography}[{\includegraphics[width=1in,height=1.25in,clip,keepaspectratio]{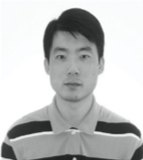}}]{Dongxiao Yu} received his BS degree in Mathematics in 2006 from Shandong University, and PhD degree in Computer Science in 2014 from The University of Hong Kong. He became an associate professor in the School of Computer Science and Technology, Huazhong University of Science and Technology, in 2016. Currently he is a professor in the School of Computer Science and Technology, Shandong University. His research interests include wireless networking, distributed computing, and graph algorithms.
\end{IEEEbiography}

\begin{IEEEbiography}[{\includegraphics[width=1in,height=1.25in,clip,keepaspectratio]{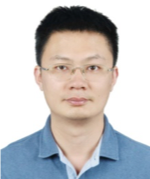}}]{Bei Gong} received his BS degree from Shandong University in 2005, and Ph.D. degree from Beijing University of Technology in 2012. Currently he is an associate professor in Computer Science at Beijing University of Technology. His research interests include trusted computing, Internet of things security, mobile Internet of things, mobile edge computing. 
\end{IEEEbiography}

\begin{IEEEbiography}[{\includegraphics[width=1in,height=1.25in,clip,keepaspectratio]{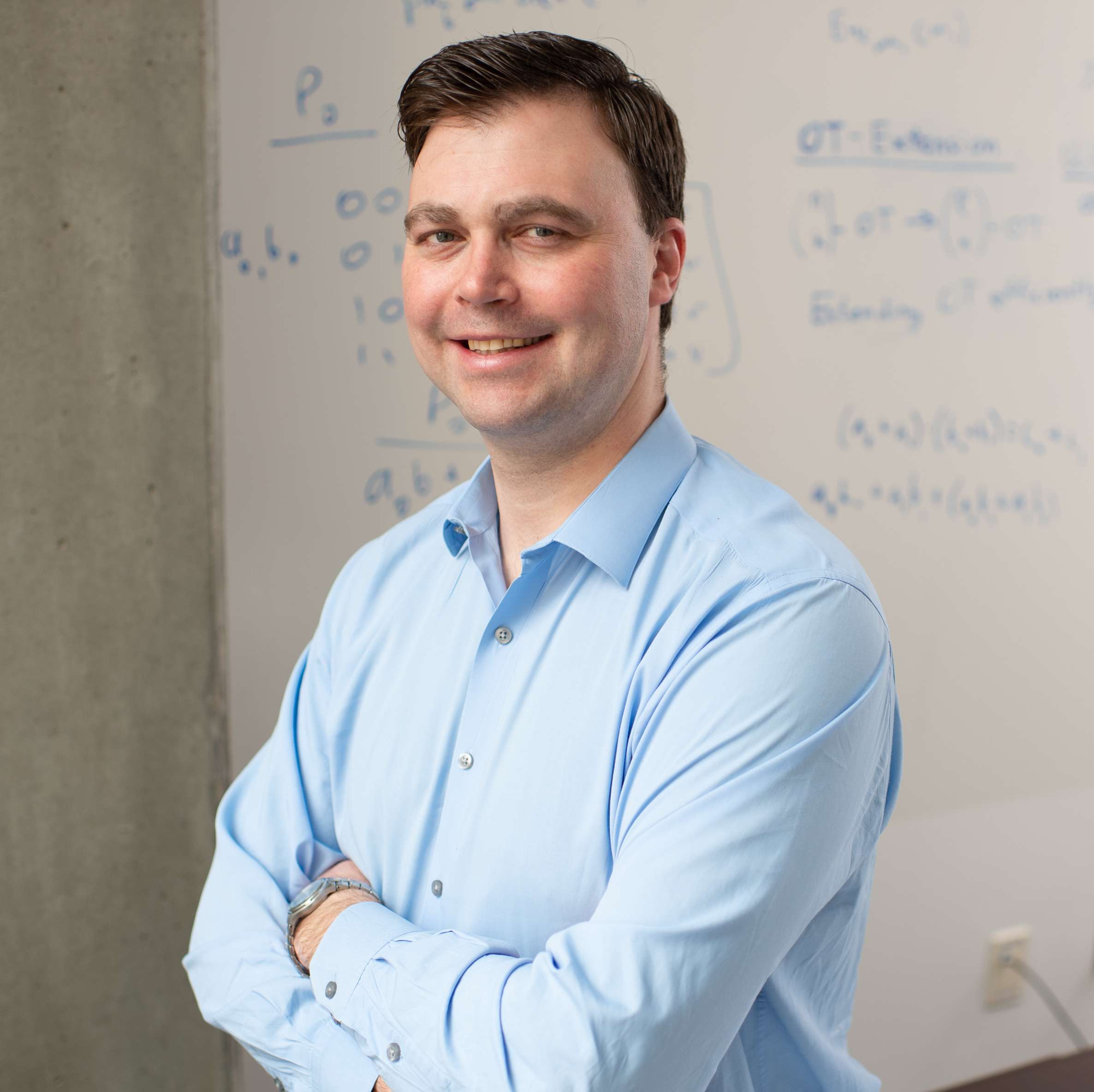}}]{Arkady Yerukhimovich} is an assistant professor at The George Washington University since Fall 2018. Prior to that, he was a research scientist in the Secure Resilient Systems and Technology Group at MIT Lincoln Laboratory where he worked on applying tools from theoretical cryptography for practical applications. He received his PhD in August 2011 under Jonathan Katz in the Computer Science department at the University of Maryland. His research aims to enable collaboration between distrusting parties.
\end{IEEEbiography}

\begin{IEEEbiography}[{\includegraphics[width=1in,height=1.25in,clip,keepaspectratio]{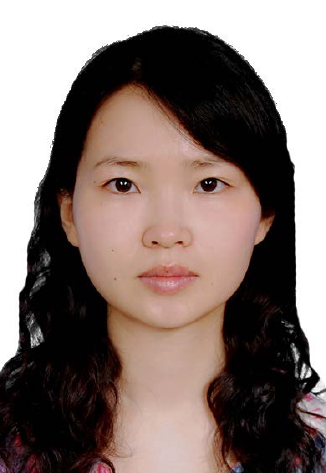}}]{Shengling Wang} is a professor in College of Information Science  and Technology, Beijing Normal University. She received her Ph.D. in 2008 from Xi’an Jiaotong University. After that, she did her postdoctoral research in the Department of Computer Science and Technology at Tsinghua University. Then she worked as a faculty member from 2010 to 2013 in the Institute of Computing Technology of the Chinese Academy of Sciences. Her research focuses on mobile/wireless networks, game theory, and crowdsourcing.
\end{IEEEbiography}

\begin{IEEEbiography}[{\includegraphics[width=1in,height=1.25in,clip,keepaspectratio]{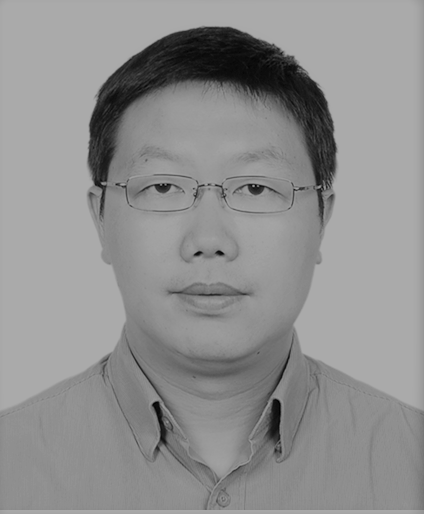}}]{Weifeng Lv} received his Ph.D. degree in computer science from Beihang University in 1998. His current research interests include massive information system, urban cognitive computing, swarm intelligence, and smart cities. He is a professor of computer science  at Beihang University. He received multiple internationally renowned awards, including the second prize of the 2016 China National Science and Technology Invention Award and the first prize of the 2010 Beijing Science and Technology Award.
\end{IEEEbiography}




\end{document}